\RequirePackage{fix-cm}
\documentclass{svjour3}                     
\smartqed  
\usepackage{mathptmx}
\usepackage{multirow, makecell}
\usepackage{booktabs}    
\usepackage{float}   
\usepackage{caption}
\usepackage{subcaption}
\PassOptionsToPackage{hyphens}{url}
\usepackage{url}
\usepackage{hyperref}
\usepackage{xcolor,colortbl}
\usepackage[graphicx]{realboxes}
\usepackage{rotating}
\usepackage[title]{appendix}

\usepackage[]{todonotes}

\usepackage{soul}
\usepackage{color}
\DeclareRobustCommand{\minor}[1]{{\sethlcolor{white}\hl{#1}}}

\DeclareRobustCommand{\checknum}[1]{{\sethlcolor{white}\hl{#1}}}

\definecolor{babyblueeyes}{rgb}{0.63, 0.79, 0.95}
\newcolumntype{a}{>{\columncolor{babyblueeyes!70}}c}
\newcolumntype{b}{>{\columncolor{babyblueeyes!70}}r}

\usepackage[framemethod=TikZ]{mdframed}
\global\mdfdefinestyle{exampledefault}{linewidth=1pt, linecolor=lightgray!80, backgroundcolor=lightgray!20, innerleftmargin=6pt, innerrightmargin=6pt, innertopmargin=4pt, innerbottommargin=4pt, nobreak=true,roundcorner=5pt}
\newenvironment{custombox}{\smallskip\begin{mdframed}[style=exampledefault]}{\end{mdframed}\smallskip}

\journalname{Empirical Software Engineering Journal}

\begin{document}

\title{
An Empirical Study on Data Leakage and Generalizability of Link Prediction Models
for Issues and Commits
}

\titlerunning{An Empirical Study on Data Leakage and Generalizability of Link Prediction Models} 

\author{Maliheh Izadi \and
        Pooya Rostami Mazrae \and
        Tom Mens \and
        Arie van Deursen
}
\authorrunning{Izadi et al.} 

\institute{M. Izadi, A. van Deursen\at
              Technical University of Delft, Netherlands \\
              \email{m.izadi@tudelft.nl, arie.vandeursen@tudelft.nl}
           \and
           P. Rostami Mazrae, T. Mens  \at
              Universit\'e de Mons, Mons, Belgium\\
              \email{pooya.rostamimazrae@umons.ac.be, tom.mens@umons.ac.be}
}

\date{Received: date / Accepted: date}

\maketitle

\begin{abstract}
Artifacts such as issues and commits interact throughout the software development cycle.
To enhance documentation and maintenance practices, developers conventionally establish links between related artifacts by manually including issue identifiers in commit messages. 
Empirical research has revealed that developers frequently overlook this practice, resulting in significant information loss. To address this issue, automatic link recovery techniques have been proposed. However, these approaches have primarily focused on improving prediction accuracy on randomly-split datasets, with limited attention given to the impact of data leakage and the generalizability of the predictive models.

The present paper introduces \textit{LinkFormer}, 
which seeks to address these limitations. 
In doing so, our approach not only preserves 
and improves the accuracy of existing predictions, 
but also enhances their alignment with real-world settings and their generalizability.
First, to better utilize contextual information for prediction,
we employ the Transformer architecture
and fine-tune multiple pre-trained models on both textual 
and metadata information of issues and commits.
Next, to gauge the effect of time on model performance, 
we employ two splitting policies during both the training and testing phases;
\textit{randomly}- and \textit{temporally}-split datasets.
Finally, in pursuit of a generic model 
that can demonstrate high performance across a range of projects, 
we undertake additional fine-tuning of LinkFormer 
within two distinct transfer-learning settings. 

Our empirical findings strongly support the notion that, 
to effectively simulate real-world scenarios,
researchers must maintain the temporal flow of data 
when training learning-based predictive models. 
Furthermore, the results demonstrate that LinkFormer 
outperforms existing state-of-the-art methodologies 
by a significant margin, 
achieving a \checknum{48\%} improvement in F1-measure within a project-based setting. 
Additionally, the performance of LinkFormer 
in the cross-project setting is comparable to its average performance 
within the project-based scenario, obtaining F1-measures of 
\checknum{88\%} and \checknum{90\%}, respectively. 
Consequently, our approach demonstrates promising results in 
applying knowledge gleaned from past projects to new, 
previously unseen contexts, 
even in the absence of historical data.
Finally, LinkFormer boasts superior runtime characteristics, 
taking less training and testing times compared to alternative methodologies. 
This efficiency can be attributed to the simplicity of our approach 
and its streamlined implementation.

\end{abstract}

\section{Introduction}\label{intro}
Software traceability refers to the ability to trace software artifacts, such as issue reports, test cases, code changes, requirements, and designs, back to their associated software entities. 
It serves to facilitate software management and maintenance tasks, including feature location, defect prediction, impact analysis, software quality measurement, and bug localization~\cite{kochhar2014potential,murali2021industry,cleland2014software}.
This work focuses on recovering links between related code changes and issue reports in a software repository. Conventionally, developers manually add an identifier to a commit to link it to its corresponding issue report(s). However, these manual links are often incomplete due to various reasons, such as a lack of motivation, guidelines, or developer negligence, necessitating the development of automatic solutions.
Two types of automatic approaches have been proposed for recovering links between commits and issue reports: heuristic-based and learning-based methods. The former has been shown to suffer from low prediction accuracy, as reported in previous studies~\cite{wu2011relink,nguyen2012multi,schermann2015discovering}. 
As a result, learning-based techniques have become more popular in recent years.
For instance, Lin et al.~\cite{lin2021traceability} 
utilized Transformers in their recent work to transfer knowledge from a related Software Engineering domain, Code Search, to the task of software traceability.
As the evaluation of the proposed approach was limited in scope (three Python projects), it should be noted that its effectiveness on a wider range of projects still needs to be investigated.
Currently, several main challenges hinder existing approaches, including low accuracy, potential data leakage, and limited generalization. However, the latter two challenges have been relatively understudied compared to the first. In the following, we elaborate on these two challenges.

In Machine Learning studies, it is customary to randomly shuffle and split data points to train and test models, assuming that the data points are independent. Furthermore, it is typically assumed that future data will be similar to the past data used to build a learning-based model.
However, the causal interaction between issues and commits in software repositories is an intrinsic characteristic of collaborative software development. This interaction is manifested through a temporal flow of information, which we believe is crucial to capture for accurate software traceability. 
For instance, when a user submits issue $I_1$ requesting a new feature, a team member may commit a code change $C_1$ to address it. Subsequently, a bug in the committed change may be reported through issue $I_2$, and the team fixes it by pushing yet another commit $C_2$, and the process continues. This example highlights the importance of incorporating the temporal flow among these artifacts into software traceability techniques. Failing to do so can result in incomplete or inaccurate links, or suggested links that are not relevant to the current changes in the software.
Consequently, randomly shuffling data can lead to the loss of the temporal flow of information in software repositories. Moreover, splitting a time-aware dataset randomly can result in data leakage between training and testing sets, potentially leading to the overestimation of a model's accuracy and its failure to generalize to real-world settings. Thus, it is essential to consider the temporal nature of the data in software repositories when training and testing predictive models for link recovery.
To the best of our knowledge, the impact of time on the performance of predictive models for recovering links among issue reports and commits has not been thoroughly investigated. 

Another important consideration for the practical applicability of these models is their ability to generalize to new projects. 
Despite extensive research in this area, most existing approaches have only been evaluated on a limited number of projects, typically around six projects per study~\cite{lin2021traceability,ruan2019deeplink,rath2018traceability,sun2017frlink,sun2017improving}. This limited evaluation may not provide a comprehensive understanding of the generalization ability of the models.
Furthermore, all the studies to date have evaluated their proposed approaches in a \textit{project-based} setting, where models are trained separately for each project and evaluated on a subset of the same project's data. This approach requires training multiple models, each specific to one project, which can be time-consuming and impractical for large-scale software engineering projects.
Therefore, there is a need for more comprehensive evaluations of link recovery models that include a larger number of projects and assess their generalization ability to new and unseen projects. 

In response to the aforementioned challenges, we present a proposed solution called \textit{LinkFormer}, a Transformer-based approach designed to enhance the accuracy and generalizability of predictive link recovery models through leveraging pre-trained language models and their fine-tuning. Furthermore, we have incorporated data and design choices in LinkFormer that reflect the realistic properties of link datasets to avoid data leakage.
LinkFormer comprises two main components that work together in unison to improve performance;

1) \textit{Data Component}: 
Our dataset is composed of \textit{True} and \textit{False links} among issue and commit pairs.
A True Link is a $\langle issue, commit \rangle$ pair 
that is manually linked by a contributor, 
while a False Link is a pair 
that are not related to each other.
These pairs are specifically curated to solve the link recovery task.
In contrast to prior research that has relied on a limited number of software projects for both training and evaluation, such as those reported in~\cite{lin2021traceability,ruan2019deeplink,rath2018traceability,sun2017frlink,sun2017improving}, we have expanded our publicly shared dataset~\cite{mazrae2021automated} by incorporating additional data from two other studies~\cite{rath2018traceability,lin2021traceability}. The updated dataset contains a combination of textual and metadata features extracted from twenty different software projects, encompassing over $233,000$ pairs of issues and commits.
More specifically, our new dataset is $7.3$ times larger than the dataset used in Rath et al.'s study~\cite{rath2018traceability} and $46.4$ times larger than the dataset employed in Lin et al.'s research~\cite{lin2021traceability}. This significant expansion of the dataset will enable more accurate and reliable evaluations of link recovery models, particularly with respect to their generalizability.
In addition to the expanded size and relevance of the dataset for the task of link recovery, our data component also offers a diverse range of projects with varying development scopes, written in multiple programming languages. Consequently, this aspect of the dataset adds to the generalizability of link recovery models, making them more effective in handling new and unseen software projects, domains, and programming languages.

2) \textit{Model Component}:
LinkFormer utilizes pre-trained Transformer-based models that come equipped with natural language text representations. To cater to the target link recovery task, we fine-tune these models using our labeled dataset. This step facilitates the model's ability to acquire task-specific patterns and fine-tune its general representations towards the specific domain of interest. In doing so, the model learns meaningful representations that capture interdependencies and similarities between issues and commits. We evaluated three architectures and identified the RoBERTa-based model~\cite{liu2019roberta} as the best-performing underlying architecture for LinkFormer.
To simulate a realistic scenario and mitigate the negative impact of data leakage, we train our model on temporally-split data in addition to randomly-split data. This reflects a real-world scenario where the training data is based on historical data, and the testing data is from a future time period. 
Furthermore, to enhance the generalizability of solutions, unlike previous approaches, we transfer knowledge within the same domain. By leveraging the existing knowledge from pre-trained models and fine-tuning processes, we equip the model to handle new and unseen projects, programming languages, and domains. 

We evaluate LinkFormer against three baselines,
including both traditional and advanced learning-based models.
DeepLink~\cite{ruan2019deeplink}
is a deep RNN model exploiting textual data,
HybridLinker~\cite{mazrae2021automated}
is an ensemble classical learner,
and T-BERT~\cite{rath2018traceability} 
is a Transformer-based model.
We evaluate our proposed LinkFormer model in four distinct settings to comprehensively assess its performance. Firstly, we train models in a project-based setup using randomly-split data. Secondly, we use temporally-split data for project-based training. These settings allow us to study the effect of different data splitting approaches on the model's performance. Thirdly, we investigate the performance of a generic model through intermediate fine-tuning. Finally, we evaluate the generalizability of our proposed models across different projects by using a cross-project setup. By exploring these different settings, we are able to gain insights into the effectiveness and generalizability of our proposed approach in various scenarios.

Based on our experiments, LinkFormer demonstrates superior performance compared to the state-of-the-art methods in both project-based and cross-project link prediction tasks. Our results also suggest that using temporal splitting is crucial to prevent data leakage and achieve a more realistic setting. Additionally, LinkFormer's performance in the intermediate fine-tuning and cross-project settings is comparable to its performance in the project-based mode, demonstrating its ability to transfer knowledge from the training projects to new, unseen test projects. Consequently, LinkFormer can be leveraged for link prediction in projects with limited or no training data.
%
%
The main contributions of this work are:
\begin{itemize}
    \item The LinkFormer dataset: 
    the largest and widest set of software projects used for the link recovery problem (Section~\ref{data_collection}). 
    \item The LinkFormer model, based on the RoBERTa architecture, is a straightforward yet effective model that demonstrates remarkable performance superiority over baselines in all cross-project and project-based settings, as described in Section~\ref{sec:model}.
    \item
    An investigation of the impact of training-testing data leakage on the link prediction task through random and temporal data splitting strategies (Sections~\ref{rq1} and \ref{rq2}).
    \item An empirical assessment of 
    the generalizability of LinkFormer (Sections~\ref{rq3} and \ref{rq4}).
    \item Openly available source code~\footnote{\url{https://github.com/MalihehIzadi/linkformer}} and dataset~\footnote{\url{https://doi.org/10.5281/zenodo.6524460}}.
\end{itemize}

\section{Background and Related Work}
To begin, we provide some background information on the problem of link prediction. 
Following that, we present an overview of the existing literature that addresses this problem.

\subsection{Software Artifacts and Linking}\label{sec:artifacts}
An Issue Tracking System (ITS) contains information pertaining to reported bugs, change requests, new features, or tasks related to the development of a project. In this context, an \emph{issue} refers to the discussion surrounding a particular topic, such as a desired change. An issue is characterized by several information fields, including a unique ID, issue type, status, priority, severity, description, creation date, updated date, and resolution information. Additionally, issues may contain comments and code snippets, which can be used to further illustrate the purpose of the issue.
A \emph{commit} refers to a collection of necessary code changes that are made within the version control system (VCS). A VCS is used for managing and documenting changes throughout the software development process. A commit is comprised of several fields, including a unique commit ID, commit message, committer ID, author ID, changed files, and their diff. The diff field encapsulates the differences between the files of the current and previous states of a repository.

Many aspects of collaborative software development exploit the links between issues and commits. To link an issue with a commit, developers who are assigned to solve the issue add the unique issue ID to the associated commit. While guidelines on how to contribute to software projects are typically available, there is no guarantee that developers will follow them in practice. Factors such as negligence, deadline pressure, and the cost and effort required to maintain well-documented records may lead to an incomplete set of links and weak traceability.
Although the guidelines for a software project may vary, a general rule for linking is to have at least one issue linked to each commit to track the reason behind each code change. 
%
Table~\ref{table:motivation_example} 
provides a sample issue and commit 
from the Apache Beam project.
This is an open-source project 
for defining and executing data processing workflows.~\footnote{\url{https://beam.apache.org/}}
The issue summary indicates a bug 
related to CI/CD tools' configuration files.~\footnote{\url{https://github.com/apache/beam/issues/23671}}
The commit is directly related to the above issue.~\footnote{\url{https://github.com/apache/beam/commit/df80a0599c4bd6c736bd68874bdb83cd02c561d1}}
Despite the clear contribution guidelines of this project,
the link between this issue and the commit 
is currently missing from the project. 
No issue ID or extensive lexical overlap 
is evident in the provided information 
which makes it hard for existing solutions 
to find the connection between them.
An automatic approach capable of extracting 
semantics and contextual information 
from textual information of such artifacts
can facilitate the process of development 
along with its management and documentation.
\begin{table}[tb]
\caption{Sample issue and commit retrieved from the Apache Beam project.}
\begin{center}
\begin{tabular}{cl}
\toprule
\textbf{Issue} & [Bug]: TPC-DS Jenkins job doesn't read all data from partitioned files\\ 
\textbf{summary} &  \\
\midrule
 & TPC-DS Jenkins job doesn't read all data from partitioned files from \\
\textbf{Issue} & gs://beam-tpcds/datasets/parquet/partitioned path.\\
\textbf{description} & It can be related to an internal structure of partitioned directories and files that are not \\
 & properly matched with provided path pattern.\\
\midrule
\midrule
\textbf{Commit} & \multirow{2}{*}{[TPC-DS] Use ``nonpartitioned'' input for Jenkins jobs}  \\
\textbf{message} & \\         
\midrule
\multirow{7}{*}{\textbf{Diff code}} & in (.test-infra/jenkins/job\textunderscore PostCommit\textunderscore Java\textunderscore Tpcds\textunderscore Dataflow.groovy)\\
    & \texttt{- `--dataDirectory=gs://beam-tpcds/datasets/parquet/partitioned'}\\
    & \texttt{+ `--dataDirectory=gs://beam-tpcds/datasets/parquet/nonpartitioned'} \\
    & in (.test-infra/jenkins/job\textunderscore PostCommit\textunderscore Java\textunderscore Tpcds\textunderscore Spark.groovy)\\
    & \texttt{- `--dataDirectory=gs://beam-tpcds/datasets/parquet/partitioned'}\\
    & \texttt{+ `--dataDirectory=gs://beam-tpcds/datasets/parquet/nonpartitioned'} \\
    & ...\\
\bottomrule
\end{tabular}
\end{center}
\label{table:motivation_example}
\end{table}

\subsection{Link Prediction Approaches}\label{related}
Link prediction between issues and commits can be classified as a subcategory of link recovery that aims to establish connections between software artifacts. A systematic literature review conducted by Aung et al.~\cite{aung2020literature} identified four major approaches in this area, namely, approaches based on information retrieval methods, heuristic, machine learning, and deep learning techniques.
In the following, we review the literature in the context of studies exploring link prediction for connecting \textit{issues} and \textit{commits}.

\paragraph{Heuristic Approaches:}
Since 2011, researchers have utilized heuristic-based approaches to recover links between issues and commits. For example, Wu et al. developed ReLink~\cite{wu2011relink}, which relies on information available in developers' changelogs. The authors utilized specific keywords such as `fixed' and `bug', or issue ID references in changelogs, and combined them with features extracted from linked issues and commits.
MLink~\cite{nguyen2012multi} is a layer-based approach that leverages both textual and code-related features for link prediction between issues and commits. On the other hand, PaLiMod~\cite{schermann2015discovering} analyzes the interlinking characteristics of commits and issues. The authors of PaLiMod introduced the concepts of Loners (one commit, one issue) and Phantoms (multiple commits, one issue) as new heuristics for more effective link recovery.
Aung et al.\cite{aung2019interactive} have introduced a tool aimed at providing better visualization for link recovery of general software artifacts. The tool is based on the method proposed by De Lucia et al.~\cite{de2012information}, which primarily utilizes heuristic and information retrieval techniques.
Due to their low precision, heuristic-based methods faced limitations in effectively recovering links between software artifacts. Consequently, learning-based approaches were proposed as an alternative.

\paragraph{Traditional Learning Approaches:}
Learning-based approaches such as RCLinker~\cite{le2015rclinker}, FRLink~\cite{sun2017frlink}, and PULink~\cite{sun2017improving} utilize binary classifiers to address this problem. RCLinker, for example, automatically generates commit messages for commits and feeds them, along with textual information from issues, to its classifier.
Sun et al.~\cite{sun2017frlink} employ a set of features, including complementary documents such as non-source documents, to train their FRLink model. In contrast, PULink~\cite{sun2017improving} uses True Links and \textit{Unlabeled} links to reduce the amount of data required for training the model.
Rath et al.~\cite{rath2018traceability} utilize a combination of process and text-related features to augment a set of existing trace links between issues and code changes. However, this approach yields highly unbalanced precision and recall scores, resulting in low F1-measure scores.
Hybrid-Linker~\cite{mazrae2021automated} presents a novel approach that involves training two separate models to learn from both textual (e.g., issue text, commit message) and non-textual data (e.g., author, committer, timestamps) to effectively recover links between issues and commits.
In 2022, Parțachi et al.~\cite{partachi2022aide} introduced an online tool called \emph{Aide-memoire}, which comprises a back-end that can be installed locally and a chrome plugin as a front-end. This tool employs Mondrian Forests as a model to predict the link between pull-requests and issues, rather than connecting issues and commits, which is the main focus of this study.
Despite the superior performance of these approaches over heuristic-based ones, the accuracy of the results can still be improved, especially when the amount of training data is limited. Additionally, relying solely on a small set of projects, platforms, and programming languages for training and evaluation purposes poses a threat to the generalizability of the findings obtained from these studies. Therefore, further research is needed to address these issues and improve the effectiveness and robustness of link prediction models.

\paragraph{Deep Learning Approaches:}
In 2019, Xie et al.~\cite{xie2019deeplink} introduced the usage of a deep learning technique, \emph{DeepLink}\cite{zhou2018deeplink}, originally designed for identity linkage, to address the issue-to-commit link recovery problem. The authors utilized class embeddings in commit codes to create a knowledge graph and utilized Continuous-Bag-of-Words and Word2Vec embeddings for commit and issue documentation. Unfortunately, the knowledge graph and replication package provided by the authors are currently inaccessible.
Similarly, Ruan et al.~\cite{ruan2019deeplink} proposed a semantically-enhanced link recovery method based on DeepLink~\cite{zhou2018deeplink}. They utilized textual data from issues and commits while omitting comments to avoid noise introduction.
However, this method requires a large amount of data and significant computational resources for training and testing which can be a limitation for some projects or organizations with limited resources or a small number of available data samples.
Most recently, Lin et al.~\cite{lin2021traceability} conducted a study that utilized Transformers to enhance traceability for software artifacts. The authors employed a pre-trained language model associated with the Code Search task and conducted experiments with three different network settings: Single, Twin, and Siamese. They fine-tuned a pre-trained language model with these three network settings and analyzed their results. Their findings suggest that the Single-BERT architecture yielded the most accurate results, whereas the Siamese-BERT network reduced execution time.
Despite the authors' success in improving the accuracy of suggested links through the use of pre-trained models, the model was trained and tested only on three projects, all of which were written in Python. Additionally, the potential impact of time and data leakage between training and testing is not addressed in these studies. Lastly, all approaches were trained and evaluated in a project-specific setting, which limits the generalizability of the provided models.

\subsection{Context and Data Characteristics}
\label{sec:data_leak}
Operational support tools as mentioned above were developed to aid in software development tasks and record an enormous volume of transactions and logs. Researchers have been scrutinizing these data to improve software development. However, it is essential to acknowledge, as emphasized by Mockus~\cite{mockus2014engineering}, that comprehending the context in which software development data is generated is crucial to developing precise methods for solving software development problems.
For instance, several pieces of research  have been devoted to highlighting potential issues associated with mining information from VCSs, ITSs, and collaborative software development platforms. 
Bird et al.~\cite{bird2009promises} analyzed Git and identified certain risks associated with mining Git repositories. 
Later, Kalliamvakou et al.~\cite{kalliamvakou2014promises} provided an overview of important factors that researchers should be aware of when mining GitHub repositories.
Despite the efforts made by previous studies to raise awareness about the characteristics of information that can be obtained from Git and GitHub repositories, many researchers still use data from these sources in an inappropriate manner due to the changeability of the fields associated with software artifacts. 
For example, 
Tu et al.~\cite{tu2018careful} analyzed $58$ studies related to detecting duplicate issues, localizing issues, and predicting issue-fix time, and revealed that $11$ of them were specifically affected by data leakage, while $44$ others were suspected to suffer from it. 

While previous research has underscored the significance of comprehending the context of a problem and the data employed in building models, these studies have not fully addressed the problems of time sensitivity and data leakage that can arise from various data splitting policies. Moreover, the impact of this issue on the specific problem of connecting related issues and commits (the main software artifacts in this research), has not been sufficiently investigated.



\section{The LinkFormer Approach}\label{approach}
\begin{figure}[tb]
    \centering
    \includegraphics[width=0.7\linewidth]{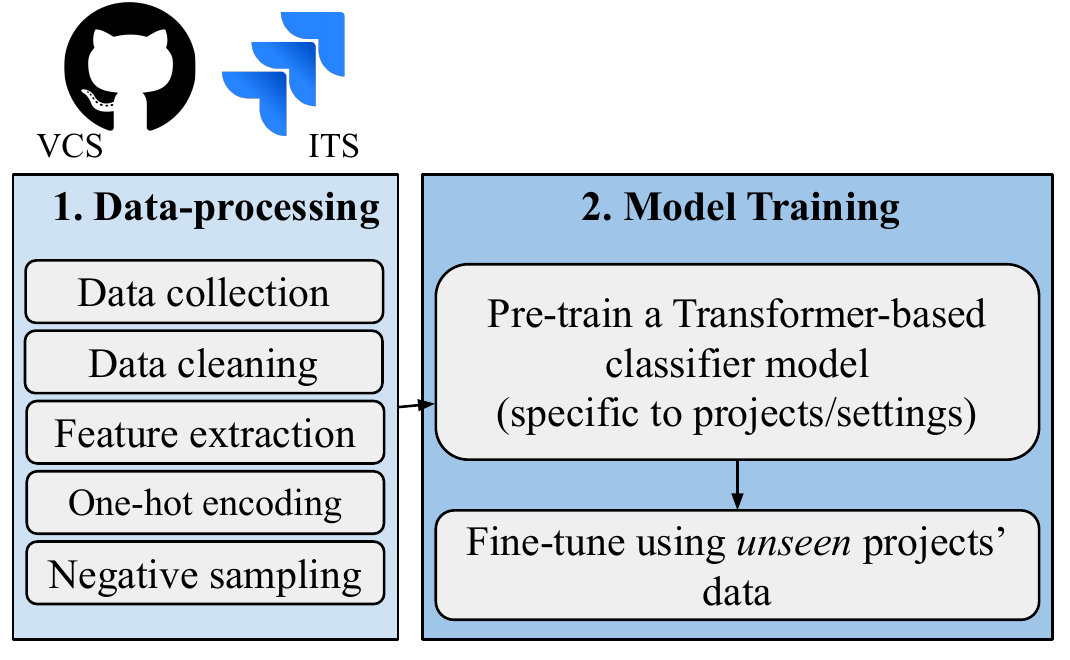}
    \caption{The workflow of LinkFormer}
    \label{fig:workflow}
\end{figure}
The general workflow of LinkFormer is presented in Figure~\ref{fig:workflow}. Our approach consists of two primary phases, namely data processing and model training. Initially, we gather and preprocess the most comprehensive set of software projects available to address the $\langle issue, commit \rangle$ pairing issue. 
Subsequently, we pre-train our Transformer model in various settings, including project-based, intermediate fine-tuning, and cross-project settings. We also utilize two distinct data splitting policies, namely random and temporal. 
In the following section, we provide more detailed information about our approach.
%

\subsection{Data Component}
\label{data_collection}
\begin{table*}[tb]
    \caption{General information of software projects included in our dataset (part 1).}
    \begin{center}
     \begin{tabular}{p{17mm}lrrrrl} 
     \toprule
     \textbf{Source} & \textbf{Project}
     & \textbf{\#Stars} & \textbf{\#Forks} & \textbf{\#Contribs} 
     & \textbf{\#True Links} & \textbf{ITS/VCS} \\ [0.15ex]
     \toprule
     \multirow{5}{*}{Rath et al.~\cite{rath2018traceability}} 
       & Maven      & 3,100 & 2,200 & 153 & 1,676 & Jira/GitHub \\
       & Pig        & 648 & 460 & 11 & 1,966 & Jira/GitHub \\
       & Derby      & 273 & 132 & 4 & 3,815 & Jira/GitHub \\
       & Drools     & 4,300 & 2,200 & 79 & 3,846 & Jira/GitHub \\
       & Infinispan & 950 & 566 & 166 & 4,640 & Jira/GitHub \\
       \midrule
      \multirow{12}{*}{Mazrae et al.~\cite{mazrae2021automated}}
       & Cassandra  & 7,200 & 3,100 & 361 & 147 & Jira/GitHub \\
       & Freemarker& 733 & 219 & 30 & 178 & Jira/GitHub \\
       & Netbeans  & 1,800 & 687 & 175 & 1,370 & Jira/GitHub \\
       & Calcite   & 3000 & 1,700 & 294 & 3,059 & Jira/GitHub \\
       & Arrow     & 9,400 & 2,300 & 719 & 5,252 & Jira/GitHub \\ 
       & Airflow   & 25,500 & 10,400 & 2,035 & 5,295 & Jira/GitHub \\ 
       & Beam      & 5,400 & 3,500 & 903 & 5,750 & Jira/GitHub \\
       & Causeway      & 684 & 281 & 47 & 8,486 & Jira/GitHub \\ 
       & Groovy    & 4,500 & 1,700 & 330 & 8,851 & Jira/GitHub \\
       & Ignite    & 4,100 & 1,800 & 277 & 9,998 & Jira/GitHub \\
       & Flink     & 18,600 & 10,600 & 1,043 & 14,472 & Jira/GitHub \\
       & Ambari    & 1,600 & 1,400 & 134 & 35,590 & Jira/GitHub \\
       \midrule
        \multirow{3}{*}{Lin et al.~\cite{lin2021traceability}}
       & Pgcli & 10,300 & 466 & 136 & 643 & GitHub/GitHub \\
       & Keras & 55000 & 19,100 & 1,014 & 719 & GitHub/GitHub \\
       & Flask & 58,600 & 15000 & 653 & 1,158 & GitHub/GitHub \\
     \bottomrule
    \end{tabular}
    \end{center}
    \label{table:projects_information_summary_1}
\end{table*}

\begin{table*}[tb]
    \caption{General information of software projects included in our dataset (part 2).}
    \begin{center}
     \begin{tabular}{p{17mm}lllp{11mm}} 
     \toprule
     \textbf{Source} & \textbf{Project}
     & \textbf{Scope/Domain} & \textbf{Top 3 Languages} 
     & \textbf{Last \newline Activity} \\ [0.15ex]
     \toprule\midrule
     \multirow{5}{*}{Rath et al.~\cite{rath2018traceability}}
       & Maven      & software project management & Java & Apr 2023\\
       & Pig        & dataflow programming env. & Java & Jan 2023\\
       & Derby      & relational database engine & Java & Aug 2019\\
       & Drools     & rule engine for Java & Java & Apr 2023\\
       & Infinispan & NoSQL cloud data store & Java & Apr 2023\\
       \hline
      \multirow{12}{*}{Mazrae et al.~\cite{mazrae2021automated}} 
       & Cassandra & NoSQL database & Java & Apr 2023 \\
       & Freemarker& template engine& Java & Apr 2023\\
       & Netbeans  & IDE & Java & Apr 2023\\
       & Calcite   & data management & Java & Apr 2023\\
       & Arrow     & data analytics app developing & C++, Java, Go & Apr 2023\\ 
       & Airflow   & monitoring workflow platform & Python & Apr 2023\\ 
       & Beam      & data processing framework & Java, Python & Apr 2023\\
       & Causeway      & domain-driven web framework & Java, Kotlin, JS & Apr 2023\\ 
       & Groovy    & programming language for JVM & Java, Groovy & Apr 2023\\
       & Ignite    & distributed database & Java, C\#, C++ & Apr 2023\\
       & Flink     & data processing framework & Java, Scala, Python & Apr 2023\\
       & Ambari    & Hadoop management tool & Java, JS, Python & Apr 2023\\
       \hline
      \multirow{3}{*}{Lin et al.~\cite{lin2021traceability}}
       & Pgcli & database CLI & Python & Feb 2023\\
       & Keras & deep learning library & Python & Apr 2023\\
       & Flask & micro-web framework & Python & Apr 2023\\
     \bottomrule
    \end{tabular}
    \end{center}
    \label{table:projects_information_summary_2}
\end{table*}

To address the issue of limited project diversity in previous studies, we aimed to provide a more comprehensive and inclusive set of projects. We accomplished this by selecting and unifying all projects utilized in three recent studies on this topic~\cite{rath2018traceability,mazrae2021automated,lin2021traceability}.
Rath et al.~\cite{rath2018traceability} investigated six projects, all in Java, while Lin et al.~\cite{lin2021traceability} conducted experiments on three projects, all in Python. 
In our previous study~\cite{mazrae2021automated}, we examined twelve projects, which encompassed eight distinct programming languages, namely Python, Java, C++, C\#, Groovy, Kotlin, JavaScript, and Scala.
\emph{Groovy} is the only common project~\cite{rath2018traceability,mazrae2021automated} in these sets.

Tables~\ref{table:projects_information_summary_1} and \ref{table:projects_information_summary_2} present a summary of the information for the projects that were included in our study. These projects exhibit a wide range of stars, forks, and True Links, and each project centers on a different Software Engineering domain. While all projects' commits were gathered from GitHub, issues were acquired from either Jira or GitHub. The dataset encompasses diverse programming languages, including Python, Java, C++, C\#, Groovy, Kotlin, JavaScript, and Scala.
It is worth noting that all projects in this dataset, except for Derby, are currently being actively maintained. Derby was last updated in August 2019.
The diversity in project scope enables us to explore our model's behavior across various domains, which contributes to the generalizability of our approach.
To complete the dataset, we gathered any missing information related to the issues and commits of all the $20$ projects in this set from GitHub or Jira.
For instance, as Lin et al.~\cite{lin2021traceability} exclusively rely on textual information from issues and commits, their dataset does not encompass non-textual features, such as metadata of authorship or timestamps of artifacts. 
Hence, for such cases, 
we developed specific crawlers that utilize the GitHub GraphQL API to extract information from the repositories and complement the dataset.
Given the variation in data sources, 
we take two steps to finalize the data for each project: 
\emph{Negative Sampling} and \emph{Data Pre-processing}.

\paragraph{Features and Data Pre-processing}
The dataset we collected comprises both textual and metadata information associated with the issues and commits. The textual data consists of (i) natural language text, namely the issue \texttt{title}, issue \texttt{description}, and \texttt{commit message}, and (ii) source code extracted from the code \texttt{diff} of a given commit.
To preprocess the textual information, we first tokenize the text and remove stop words.
The metadata associated with issues and commits encompasses information such as the \texttt{author}, \texttt{committer}, \texttt{timestamps}, as well as categorical data comprising the issue type and commit status.
It is worth noting that each project on a hosting platform may define a unique set of values for the issue type.
To standardize the various values, we categorize the issue types into three main categories: \texttt{task}, \texttt{feature}, and \texttt{bug}.
Likewise, we categorize the commit status types into two main categories, namely \texttt{closed} and \texttt{resolved}. We then encode these categorical features using the one-hot encoding method.
Git commits include two timestamps: the author date and the committer date. The author date indicates when the commit was originally made, while the committer date represents the last time the commit was modified. These two timestamps are usually the same, but some actions, such as rebasing, can change the committer date. Flint et al.~\cite{flint2021escaping} recommend using the author date over the committer date, and since these two timestamps are correlated, we only use the author date in our study.
In order to process the timestamps associated with issues and commits, we removed the hour value and kept only the date. This was done to remove unnecessary granularity in the data, as the specific time of day is not relevant for our analysis. By keeping only the date, we were able to capture the general trends and patterns in the data.
%
The final list of features includes
\texttt{creator}, \texttt{author}, \texttt{committer}, 
\texttt{closed}, \texttt{resolved},
\texttt{bug}, \texttt{feature}, \texttt{task},
\texttt{committed\_time}, \texttt{authored\_time}, 
\texttt{created\_date}, and \texttt{updated\_date}.

\paragraph{Negative Sampling}
Negative sampling is a technique used to generate negative samples for a binary classification problem where the positive samples are already known. In our case, the positive samples are the True Links that we extract directly from the repositories through identifying documented issue IDs in associated commit titles or bodies. 
To generate negative samples, 
we use a standard negative sampling technique 
proposed in the literature~\cite{mazrae2021automated,sun2017frlink,ruan2019deeplink}.
This technique connects the previously paired issues 
from the True Link set to unrelated commits (not currently in the True Link set). 
To be more precise, 
we randomly sample pairs of issues and commits that are not directly connected by developers, and hence, are not True Links.
However, the number of False Links generated through this approach can be much larger than the number of True Links, which can create a class imbalance problem during training. To mitigate this, we can use techniques such as oversampling the minority class (True Links) or undersampling the majority class (False Links) to balance the classes.
To address this problem, we apply the following constraints.
(1) The issue and commit should belong to the same project,
(2) The issue and commit should have occurred within a specified time window to ensure that they are not too far apart in time. We set the temporal gap threshold to one week in our experiments, which is a common practice in the literature~\cite{mazrae2021automated,sun2017frlink,ruan2019deeplink}.
(3) The commit should not reference the issue in its title or body, to ensure that it is not a True Link.
(4) The issue and commit should not have been created by the same author, to ensure that they are not too closely related.
This process significantly reduces the number of False Links, 
which enables us to balance the dataset and train models effectively.
Lastly, to provide a balanced dataset 
for training purposes,
we randomly select an equal number of 
False Links as True Links for each project.

\subsection{Model Component}
\label{sec:model}
\begin{figure}[tb]
    \centering
    \includegraphics[width=0.85\linewidth]{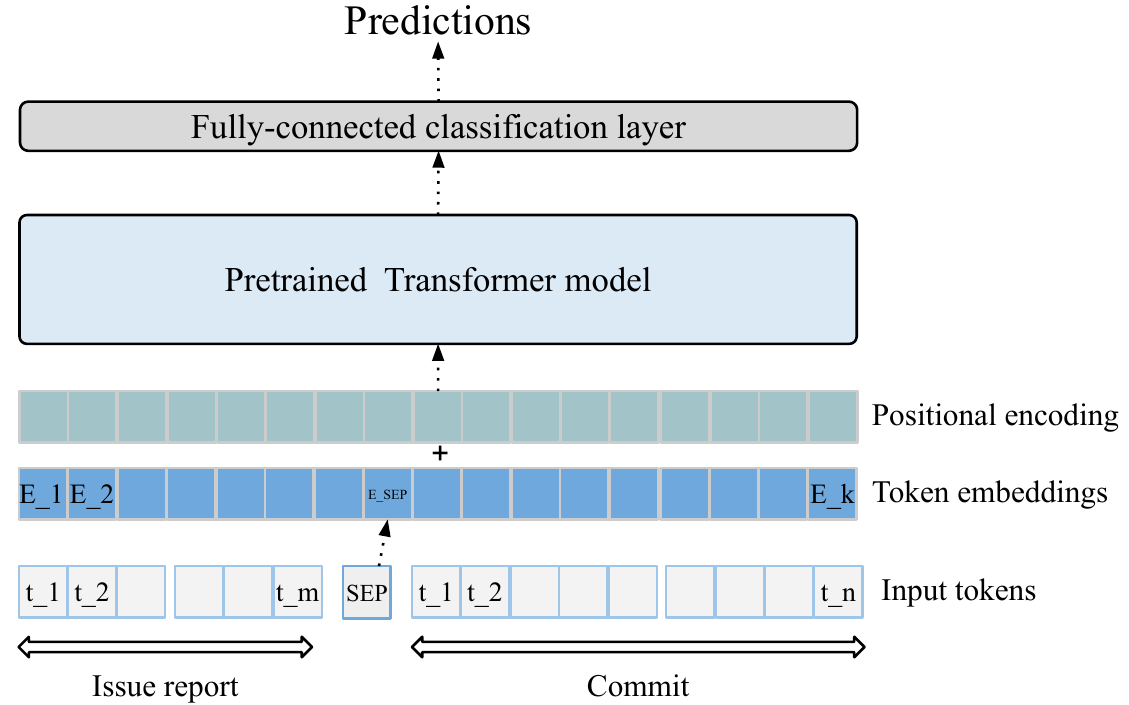}
    \caption{Training LinkFormer}
    \label{fig:approach}
    \vspace{-6mm}
\end{figure}

Over the last few years, Transformer models have made a significant impact in the field of Natural Language Processing (NLP). Pre-trained models such as Bidirectional Encoder Representations from Transformers (BERT)~\cite{devlin2018bert} have emerged as a powerful tool that can be fine-tuned by incorporating additional output layers to create state-of-the-art models without requiring significant task-specific architecture modifications. Recently, these NLP-based models have been increasingly employed for various tasks related to software engineering and source code analysis~\cite{izadi2022codefill,izadi2021topic,izadi2022predicting,al2023extending}.
In this work, we harness the capabilities of pre-trained models and fine-tune them on our dataset to achieve better performance across a wide range or projects, even in cases where little or no historical information is available.
Figure~\ref{fig:approach} illustrates the primary architecture of LinkFormer.
In the following, we elaborate on the architecture of our proposed approach.

\subsubsection{Pre-trained Model}
The Transformer architecture and the self-attention mechanism have facilitated noteworthy progress in recent years. The Transformer is a sequence-to-sequence model that can transform a given sequence of elements into another form~\cite{vaswani2017attention}.
A basic Transformer architecture comprises two main components: an encoder and a decoder. The integration of attention enables Transformer models to selectively focus on specific portions of input sequences, resulting in the production of more pertinent outputs~\cite{luong2015effective}. 
Conventional models that rely on extracting exact word matches often face challenges in identifying semantically similar text with a common context. By adopting the Transformer architecture, our proposed model, LinkFormer, can extract more contextual information from the data, thereby reducing the semantic gap between issue reports and commit messages, and ultimately resulting in higher accuracy. The semantic gap refers to the difficulty in detecting semantically similar $\langle issue, commit \rangle$ pairs that are written in different lexicons.
Additionally, when compared to conventional deep learning models such as RNNs and LSTMs, Transformers have exhibited superior performance across multiple NLP tasks~\cite{vaswani2017attention}. 
Furthermore, Transformers are more parallelizable and necessitate considerably less training time compared to RNNs.

Numerous architectures have been proposed based on this concept.
The BERT model~\cite{devlin2018bert} was originally introduced to pre-train deep bidirectional representations via two pre-training tasks, Masked Language Modeling (MLM) and Next Sentence Prediction (NSP).
DistilBERT~\cite{sanh2019distilbert} is a compact general-purpose language model that integrates both language modeling and distillation. In comparison to BERT, DistilBERT is $40\%$ smaller in size and $60\%$ faster.
RoBERTa (Robustly-optimized BERT approach)~\cite{liu2019roberta} adapts BERT's MLM task by leveraging a dynamic masking technique and eliminates its NSP task.
ALBERT~\cite{lan2019albert} concentrates on reducing the memory consumption while enhancing the training speed. ALBERT attains an $80\%$ reduction in the parameters of the projection block, with only a slight decrease in performance compared to BERT.
Given that pre-training from scratch is highly expensive,
we employ the existing pre-trained models in the NLP domain, 
and build upon their knowledge 
to solve our downstream Software Engineering task: 
the link recovery problem.
In this work, 
we employ the above-mentioned three recent 
and high-performing Transformer models; 
RoBERTa~\cite{liu2019roberta}, 
DistilBERT~\cite{sanh2019distilbert},
and AlBERT~\cite{lan2019albert}
as part of our proposed approach.
We have selected these variants 
with three aspects in mind:
high accuracy, reasonable model size, and short inference time.

\subsubsection{Fine-tuning}
Fine-tuning the pre-trained models on our dataset 
enables us to adjust the parameters of LinkFormer to the link prediction task.
Our model takes pre-processed $\langle issue, commit \rangle$ pairs as input sequences, along with their corresponding labels (True or False Link) to indicate whether each pair is related or not.
To generate the input vectors for each issue and commit, we concatenate the summary and metadata for the issue and the message and metadata for the commit. This results in an issue vector and a commit vector, respectively.
Note that our approach is programming language-agnostic, as we do not heavily preprocess text or process source code. This means that LinkFormer can be applied to any software project without requiring extensive modifications or language-specific preprocessing steps.
We incorporate a fully-connected classification layer over the underlying pre-trained model to achieve fine-tuning to adapt to our specific task of link recovery.
In the context of LinkFormer, 
probability distributions are generated using the sigmoid activation function implemented in the output layer.

\section{Experiment Design}\label{design}
This section commences with an outline of our Research Questions (RQ). We then present the selected baselines and the evaluation metrics employed to measure the effectiveness of our model compared to baselines. Finally, we provide our configuration and implementation details for reuse and replication purposes.

\subsection{Research Questions}
\label{rq}
To assess the efficacy of our proposed approach in project-based and cross-project settings, we define four RQs. These questions serve as the basis for measuring the effectiveness of our approach.

\begin{itemize}
    \item \textbf{RQ1 seeks to determine the accuracy of LinkFormer in predicting links in a project-based setting, using randomly-split data}.
This is the basic setting commonly used in literature in which the models are trained and evaluated on each project separately. The dataset samples are randomly shuffled, with 80\% of the data used for training and the remaining 20\% used for validation and testing (each use 10\% of data).
\item \textbf{RQ2 aims to determine the accuracy of LinkFormer in recovering links in a project-based setting using \emph{temporally}-split data}.
To account for potential accidental data leaks arising from time dependencies, the models are trained on temporally-sorted data. Specifically, the first 80\% of each project's data is used for training, and the remaining 20\% is used for validation and testing (each use 10\% of data).
\item \textbf{RQ3 seeks to evaluate the accuracy of LinkFormer when employing \emph{intermediate} fine-tuning}. 
To address the issue of projects with insufficient data, we train a generic model capable of recovering links for such projects.
To train such a model capable of recovering links for projects with insufficient data, we first pre-train LinkFormer on 80\% of the data from 16 projects. Next, we update the model's parameters through an additional round of intermediate fine-tuning, followed by validation using 80\% and 10\% of the data from each of the four test projects. 
This intermediate fine-tuning phase allows the model to learn from a portion of historical data from the four test projects while also embedding a body of knowledge from the initial 16 projects (achieved through transfer learning via pre-trained models). 
Finally, we evaluate the model's performance on the remaining 10\% of the data from the four projects in the test set.
\item \textbf{RQ4 aims to evaluate the accuracy of LinkFormer in a \emph{cross-project} setting}. 
In the second mode of transfer learning, we assess the model's performance on projects for which it has not seen \emph{any} historical information. This is achieved by using the pre-trained model, fine-tuned on the data from 16 projects, directly for evaluation on the test set of four unseen test projects. We employ a cross-fold validation setting, dividing the dataset into five folds, each containing four projects. In each experiment, we use four folds for fine-tuning and one fold for testing, rotating the folds until all 20 projects are assessed in the testing set. 
It is worth noting that experiments for RQ3 and RQ4 are both based on temporally-split data.
\end{itemize}


\subsection{Baseline Approaches}\label{baselines}
To conduct a comprehensive evaluation, we consider three recent approaches for link recovery as our baselines~\cite{ruan2019deeplink,mazrae2021automated,lin2021traceability}.
Hybrid-Linker~\cite{mazrae2021automated} 
is based on a traditional supervised classifier, namely decision trees.
Ruan et al.~\cite{ruan2019deeplink} 
employ an RNN,
and T-BERT~\cite{lin2021traceability} 
uses a state-of-the-art Transformer model for NLP.
Hybrid-Linker~\cite{mazrae2021automated} 
extracts different features from data 
and separates them into a textual and non-textual channels.
The textual data of projects is fed into a Gradient Boosting model, while the non-textual data is used to train an ensemble model based on Gradient Boosting and XGBoost techniques.
The final result in Hybrid-Linker is obtained by combining the decisions of these separate models.
Ruan et al.~\cite{ruan2019deeplink} 
use a word embedding module, an RNN module, and a similarity module as part of their link recovery method. 
Lin et al.~\cite{lin2021traceability} 
generate trace links between source code and natural language artifacts using various Transformer-based architectures, namely Single-BERT, Twin-BERT, and Siamese-BERT.
Based on the results reported by the authors, the Single-BERT architecture generates the most accurate links. Therefore, in our experiments, we use their best-performing model, Single-BERT~\cite{lin2021traceability}.

It is important to note that all of these baseline approaches shuffle their data randomly and then split their training and test set. As a result, they do not report the performance of their models on temporally-sorted data.
Additionally, HybridLinker and DeepLink are not designed to operate in a cross-project setting. They are limited to recovering links within a single project.
Finally, none of the baselines evaluate 
the performance of their proposed model 
on \emph{unseen} projects.
Hence, for addressing RQ1 and RQ2, we have incorporated DeepLink, HybridLinker, and T-BERT as our baseline approaches. However, in the case of RQ3 and RQ4, where transfer learning mode is involved, T-BERT is the only viable option as it is the sole approach that relies on pre-trained models. 

\subsection{Evaluation Metrics}\label{sec:eval_metrics}
In the assessment of link prediction models, precision, recall, and F1-measure are the three standard measures used. A model with low precision generates numerous false positive predictions, requiring manual verification of the suggested links by a human expert. Similarly, a link recovery method with low recall is unable to retrieve the actual linked $\langle issue, commit \rangle$ pairs in the data, resulting in a significant decline in the model's overall performance. Therefore, a good predictive model should exhibit high precision and recall. 
Hence, we primarily compare the performance of models based on their F1-measure, which is the harmonic mean of precision and recall:
The following equations are employed to calculate the aforementioned metrics:
\begin{equation}
    Precision=\frac{TP}{TP + FP}
\end{equation}
\begin{equation}
    recall=\frac{TP}{TP + FN}
\end{equation}
\begin{equation}\label{eq:f1}
    F1=\frac{2 \times precision \times recall}{precision + recall}
\end{equation}
Here, TP, FP, TN, and FN represent True Positive, False Positive, True Negative, and False Negative predictions, respectively.

\subsection{Implementation and Configuration}
\label{sec:config}
For pre-processing, we utilize the Pandas Python library~\cite{mckinney2011pandas}. 
In the case of training traditional classifiers, we employ the Sci-Kit Learn library.~\footnote{\url{https://scikit-learn.org}} 
Furthermore, we use the HuggingFace~\footnote{\url{https://huggingface.co}} 
and SimpleTransformers~\footnote{\url{https://gitbub.com/ThilinaRajapakse/simpletransformers}} 
libraries 
for the implementation of LinkFormer. 
We set the learning rate to $3 * 10^{-5}$, 
the number of fine-tuning epochs to $6$, 
the maximum input length to $512$, 
and the batch size to $32$.
To ensure consistency and accuracy in our experimentation, we utilized the provided replication package for all the baselines.
In the project-based setting, we split the data of each project into training, validation, and testing sets, using an 80/10/10 percent ratio, respectively.
In the fine-tuning setting, we divide the projects into five folds. Each time, we utilize four folds (16 projects) for fine-tuning, and one fold (4 projects) for testing.
All our experiments have been conducted on a server
equipped with one GeForce RTX 2080 GPU,
an AMD Ryzen Threadripper 1920X CPU 
with $12$ core processors, and $64$G of RAM.
Our source code and dataset are available online for public access.~\footnote{\url{https://github.com/MalihehIzadi/linkformer}}

\section{Results}\label{results}
In this section, we present the outcomes of our experimentation. We compare our proposed model with three baselines, which include a traditional learning-based model~\cite{mazrae2021automated}, 
a deep learning model (RNN)~\cite{ruan2019deeplink}, 
and a Transformer-based model~\cite{lin2021traceability}.
We report the results based on two data splitting choices, which are random and temporal. Additionally, we fine-tune and evaluate the models in both project-based and transfer learning modes.

\subsection{RQ1: Project-based Training with Randomly-split Data}\label{rq1}
\begin{table*}[tb]
    \caption{Project-based training and assessment (RQ1: randomly-split datasets)}
    \begin{center}
     \begin{tabular}{l|r|r|r|r} 
     \toprule
     \textbf{Project}
     & Deeplink & Hybrid-Linker & T-BERT & LinkFormer \\
     & \cite{ruan2019deeplink} & \cite{mazrae2021automated} & \cite{lin2021traceability} & (Proposed)\\[0.15ex]
     \midrule
       Maven    & 0.66 & 0.81  & 0.90 & \textbf{1.00} \\
       Pig    & 0.66 & 0.74  & \textbf{1.00} & \textbf{1.00} \\
       Derby    & 0.63 & 0.85  & \textbf{1.00} & \textbf{1.00} \\
       Drools    & 0.64 & 0.84  & \textbf{1.00} & \textbf{1.00} \\
       Infinispan    & 0.55 & 0.87  & \textbf{1.00} & \textbf{1.00} \\
       \hline
       Cassandra    & 0.80 & \textbf{0.91} & NA & 0.90 \\
       Freemarker    & 0.89 & 0.87  & 0.11 & \textbf{0.97} \\
       Netbeans    & 0.59 & 0.88  & 0.65 & \textbf{0.94} \\
       Calcite    & 0.66 & 0.87  & 0.94 & \textbf{0.99} \\
       Arrow   & 0.43 & 0.84  & 0.94 & \textbf{0.97} \\ 
       Airflow   & 0.45 & 0.86  & 0.97 & \textbf{0.98} \\ 
       Beam   & 0.66 & 0.86  & 0.70 & \textbf{0.95} \\
       Causeway   & 0.79 & 0.88  & \textbf{0.92} & \textbf{0.92}\\ 
       Groovy   & 0.64 & 0.88  & \textbf{0.97} & \textbf{0.96} \\
       Ignite   & 0.64 & 0.90  & 0.81 & \textbf{0.94} \\
       Flink   & 0.72 & 0.91  & 0.94 & \textbf{0.97} \\
       Ambari   & 0.82 & 0.96  & \textbf{1.00} & \textbf{0.99} \\
       \hline
       Pgcli   & 0.49 & 0.67  & 0.89 & \textbf{0.95} \\
       Keras   & 0.35 & 0.75  & 0.97 & \textbf{0.99} \\
       Flask   & 0.42 & 0.68  & 0.74 & \textbf{0.91} \\
       \hline
       \rowcolor{babyblueeyes!70}
       Average score & 0.62 & 0.84 &  0.86 & \textbf{0.97} \\
       \hline
       Standard deviation & 0.15 & 0.07 &  0.21 & \textbf{0.03} \\
     \bottomrule
    \end{tabular}
    \end{center}
    \label{table:random_split_results}
    \vspace{-6mm}
\end{table*}

Table~\ref{table:random_split_results} displays the outcomes of the first experiment, in which we trained each model per project (project-based mode). Here, the data of each project is randomly partitioned into training, validation, and testing sets.
The last two rows of this table present 
the mean average score of a model 
and its standard deviation
over all projects.
To prevent the table from becoming too cluttered due to space constraints, for RQ1 and RQ2, we report only the F1-measure of all approaches in this section. For precision and recall scores of each approach please refer to the Appendix~\ref{appendix:precision_recall}.

As mentioned before, we consider and evaluate multiple architectures to identify the best-performing one for our proposed model, LinkFormer. Based on our experiments, we found that the RoBERTa-based model~\cite{liu2019roberta} provided the best results among the evaluated architectures. Therefore, we report LinkFormer's performance based on this architecture.
Based on the results of the first experiment, 
we concluded that LinkFormer outperforms the three baselines, 
namely, DeepLink, HybridLinker, and T-BERT by
\checknum{56\%}, \checknum{15\%}, and \checknum{13\%}, respectively.
Note that all the results obtained in the randomly-split case are consistent with previous studies. However, for the Cassandra project, T-BERT's replication package encounters errors and does not provide predictions. Hence, the results for this project/model are marked as Not Available (NA). This may be due to the fact that this project has the lowest number of True Links in the dataset.
%

\subsection{RQ2: Project-based Training with Temporal Data Split}\label{rq2}
\begin{table*}[tb]
    \caption{Project-based training and assessment (RQ2:  temporally-split datasets)}
    \begin{center}
     \begin{tabular}{l|r|r|r|r}
     \toprule
     \textbf{Project}
     & Deeplink & Hybrid-Linker & T-BERT & LinkFormer  \\
     & \cite{ruan2019deeplink} & \cite{mazrae2021automated} & \cite{lin2021traceability} & (Proposed) \\[0.15ex]
     \midrule
       Maven    & 0.79 & 0.85 & 0.78 & \textbf{0.97} \\
       Pig    & 0.83 & 0.76 & 0.75 & \textbf{0.99}\\
       Derby    & 0.47 & 0.86 & 0.54 & \textbf{1.00} \\
       Drools    & 0.47 & 0.74 & 0.94 & \textbf{0.99} \\
       Infinispan    & 0.57 & 0.85 & 0.43 & \textbf{0.99} \\
       \hline
       Cassandra    & 0.75 & 0.63 & NA & \textbf{0.77} \\
       Freemarker    & 0.13 & 0.42 & 0.16 & \textbf{0.71} \\
       Netbeans    & 0.09 & 0.44 & 0.81 & \textbf{0.90} \\
       Calcite    & 0.44 & 0.60 & 0.91 & \textbf{0.97} \\
       Arrow   & 0.43 & 0.69 & 0.91 & \textbf{0.97}\\
       Airflow   & 0.64 & 0.54 & 0.60 & \textbf{0.98} \\
       Beam   & 0.54 & 0.58 & 0.50 & \textbf{0.75} \\
       Causeway   & 0.54 & 0.40 & 0.12 & \textbf{0.70} \\
       Groovy   & 0.60 & 0.54 & 0.50 & \textbf{0.91} \\
       Ignite   & 0.52 & 0.78 & 0.52 & \textbf{0.94} \\
       Flink   & 0.65 & 0.68 & 0.73 & \textbf{0.94} \\
       Ambari   & 0.67 & 0.84 & 0.37 & \textbf{0.94} \\
       \hline
       Pgcli   & 0.62 & 0.72 & 0.70 & \textbf{0.75} \\
       Keras   & 0.50 & 0.78 & 0.82 & \textbf{0.90} \\
       Flask   & 0.57 & 0.62 & 0.50 & \textbf{0.93} \\
       \hline
       \rowcolor{babyblueeyes!70}
       Average score  & 0.54 & 0.67 & 0.61 & \textbf{0.90} \\
       \hline
       Standard deviation   & 0.18 & 0.14 & 0.23 & \textbf{0.10} \\
     \bottomrule
    \end{tabular}
    \end{center}
    \label{table:temporal_split_results}
    \vspace{-6mm}
\end{table*}

Table~\ref{table:temporal_split_results} 
presents the results of the second experiment.
Contrary to the previous setup, 
we split the data of each project temporally. 
We use the first $80\%$ and $10\%$ for training and validation, 
and the last $10\%$ of data (newest data) for testing the models.
Based on the F1-measure of predictions in the second setting, the results indicate that the performance of LinkFormer surpasses that of all baselines. 
More specifically,
LinkFormer outperforms DeepLink, HybridLinker, and T-BERT by
\checknum{67\%}, \checknum{34\%}, and \checknum{48\%}, respectively.
Comparing the results from the first two RQs,
it is evident that all approaches (baselines as well as the proposed approach)
perform weaker 
on the temporally-sorted data.
However, this drop in performance varies among models 
($7\%$ to $25\%$ decline in F1-measure).
For instance, T-BERT suffers the most, 
while LinkFormer's F1-measure is the least affected.

%

An inspection of the results per project reveals that the performance of a given model is influenced by the unique characteristics of each project. Factors that may impact the performance of a model include the project's programming language, scope, as well as the quality and quantity of its data.
While there is a notable variance among the baselines' performance 
across different projects in both data splitting settings, 
LinkFormer demonstrates lower standard deviation compared to all baselines. 
As an illustration, in the temporal splitting mode, LinkFormer achieves an average F1-measure of $90\%$, with a standard deviation of $10\%$. In contrast, T-BERT exhibits the highest standard deviation in both splitting cases, indicating a wide range of performance scales across different projects (on average 61\% F1-measure with a standard deviation of $23\%$). 
These findings suggest that LinkFormer is a more reliable model, as it consistently performs well across different projects, whereas T-BERT's performance may be less predictable and more variable.

%
Another factor that should be taken into account is the variability among data sources. For instance, the five projects analyzed in Rath et al.'s study~\cite{rath2018traceability} (the first five projects listed in Table~\ref{table:temporal_split_results}) occasionally contain tags, which can impact the performance of learning-based models on these projects, resulting in generally higher scores compared to other projects. This indicates that the quality and structure of the data can significantly influence the performance of models, and thus should be considered when selecting an appropriate model.

\begin{custombox}
\textbf{Insights based on the splitting policy:}
A comparison of the results obtained from the first two RQs clearly demonstrates that all approaches, including the proposed method and baselines exploit training-testing data leakage in randomly split data. Consequently, to obtain a realistic assessment of the model's performance in practical settings where future data is unavailable, it is crucial to employ a temporal split strategy for the training, validation, and test data, as adopted in RQ2. 
By using temporal splitting, it is possible to achieve a more accurate evaluation of the model's performance under realistic conditions, free from data leakage, and better inform the selection of appropriate models for software engineering tasks.
\end{custombox}


\subsection{RQ3: Generic Models via Intermediate Fine-tuning}\label{rq3}
\begin{table}[tb]
    \caption{Transfer learning (RQ3: intermediate fine-tuning)}
    \begin{center}
     \begin{tabular}{c|cca|cca} 
     \toprule
     \multirow{2}{*}{\textbf{Project}} 
     & \multicolumn{3}{c}{T-BERT*~\cite{lin2021traceability}} & \multicolumn{3}{c}{LinkFormer (Proposed)}\\\cmidrule{2-7}
     & P & R & F1
     & P & R & F1 \\[0.15ex]
     \midrule
       Flask  & \textbf{0.94}&0.94&\textbf{0.94} & 0.82&\textbf{1.00}&0.90\\
       Beam & 0.51&\textbf{1.00}&0.67 & \textbf{0.93}&0.63&\textbf{0.75}\\
       Maven & 0.76&0.94&0.84 & \textbf{0.98}&\textbf{0.99}&\textbf{0.99}\\
       Netbeans & \textbf{1.00}&0.82&0.90 & 0.92&\textbf{0.92}&\textbf{0.92} \\
       \midrule
       Ambari & 0.71&0.88&0.78 & \textbf{0.97}&\textbf{0.93}&\textbf{0.95} \\
       Freemarker & 0.41&\textbf{0.87}&0.55 & \textbf{1.00}&0.59&\textbf{0.74} \\
       Cassandra & NA&NA&NA & \textbf{1.00}&\textbf{0.79}&\textbf{0.88}\\
       Calcite & 0.87&0.82&0.84 & \textbf{0.98}&\textbf{0.97}&\textbf{0.97}\\
       \midrule
       Flink & 0.39&\textbf{0.94}&0.55 & \textbf{0.96}&0.90&\textbf{0.93}\\
       Pig & 0.70&0.92&0.79 & \textbf{0.97}&\textbf{1.00}&\textbf{0.99}\\
       Derby & 0.76&0.94&0.84 & \textbf{1.00}&\textbf{0.98}&\textbf{0.99}\\
       Drools & 0.70&0.95&0.80 & \textbf{1.00}&\textbf{0.99}&\textbf{0.99}\\
       \midrule
       Causeway & 0.37&0.17&0.23 & \textbf{0.77}&\textbf{0.64}&\textbf{0.70}\\
       Arrow & 0.89&0.95&0.91 & \textbf{0.97}&\textbf{0.98}&\textbf{0.97}\\
       Airflow & 0.81&0.76&0.78 & \textbf{0.93}&\textbf{0.97}&\textbf{0.95}\\
       Infinispan & 0.72&0.94&0.81 & \textbf{0.97}&\textbf{1.00}&\textbf{0.99}\\
       \midrule
       Groovy & 0.27&0.76&0.39 & \textbf{0.90}&\textbf{0.87}&\textbf{0.88}\\
       Ignite & 0.70&\textbf{1.00}&0.82 & \textbf{0.96}&0.89&\textbf{0.92}\\
       Pgcli & 0.41&0.78&0.53 & \textbf{0.74}&\textbf{0.91}&\textbf{0.82}\\
       Keras & \textbf{0.78}&0.81&0.79 & 0.73&\textbf{0.96}&\textbf{0.83}\\
       \hline
       \rowcolor{babyblueeyes!70}
       Avg. & 0.66&0.85&0.72 & \textbf{0.92}&\textbf{0.89}&\textbf{0.90} \\
       \hline
       Standard deviation & 0.21&0.18&0.18 & \textbf{0.08}&\textbf{0.13}&\textbf{0.09} \\
     \bottomrule
    \end{tabular}
    \end{center}
    \label{table:fine_tuning_results}
    \vspace{-6mm}
\end{table}

In this section, we present the findings of our experiments on the two pre-trained models, namely T-BERT and LinkFormer, to evaluate the impact of transfer learning when used in fine-tuning mode. The objective of this analysis is to assess the effectiveness of pre-trained models in adapting to new software engineering tasks with limited labeled data, and to investigate whether they can outperform traditional approaches that do not utilize transfer learning.
As previously stated, our study incorporates the baseline established through the pre-training approach (T-BERT~\cite{lin2021traceability}) for the next cases. The other two baselines, HybridLinker~\cite{mazrae2021automated} and DeepLink~\cite{ruan2019deeplink}, are excluded due to their original proposal and design that caters to project-based training and testing. Consequently, they exhibit inferior performance when dealing with unfamiliar projects, as they do not incorporate the transfer learning technique.
Table~\ref{table:fine_tuning_results} 
provides the results of this experiment.
Each fold contains $16$ projects for training 
and four projects for testing. 
We report the evaluation results 
for these four projects in each fold, 
with each fold separated by horizontal lines for clarity.

It is worth noting that T-BERT, being a pre-trained model, has undergone fine-tuning with supplementary data, i.e., the CodeSearchNet dataset, as per the authors' design decisions. Nonetheless, to avoid placing T-BERT in a disadvantageous position when addressing RQ3, we fine-tune this baseline by utilizing 80\% of our data from each of the four test projects. This process mirrors that of our proposed approach, with the exception that our models do not possess the knowledge acquired by T-BERT from the CodeSearchNet dataset.
Notwithstanding, LinkFormer surpasses T-BERT in performance by \checknum{$25$\%} even in this scenario, indicating that our approach and dataset is more suitable for fine-tuning models for this particular task.
Moreover, inspecting the results per project,
for $19$ out of $20$ projects, 
LinkFormer outperforms T-BERT by large margins excpet for the Flask project.
%
%
Lastly, it is worth noting that once again 
LinkFormer displays a reduced standard deviation in comparison to T-BERT, 
signifying a more consistent performance across all projects and the five folds.

\subsection{RQ4: Generic Models via Cross-project Training}\label{rq4}
\begin{table}[tb]
    \caption{Transfer learning  (RQ4: cross-project mode)}
    \label{table:cross_project_results}
    \begin{center}
     \begin{tabular}{c|cca|cca} 
     \toprule
     \multirow{2}{*}{\textbf{Project}} 
     & \multicolumn{3}{c}{T-BERT*~\cite{lin2021traceability}} & \multicolumn{3}{c}{LinkFormer (Proposed)}\\\cmidrule{2-7}
     & P & R & F1
     & P & R & F1\\[0.15ex]
     \midrule
      Maven & 0.81&\textbf{1.00}&0.89 & \textbf{0.98}&0.99&\textbf{0.99} \\
      Netbeans & 0.83&0.88&0.85 & \textbf{0.90}&\textbf{0.93}&\textbf{0.91} \\
      Flask  & \textbf{0.85}&0.35&0.50 & 0.72&\textbf{1.00}&\textbf{0.84}\\
      Beam & \textbf{1.00}&0.29&0.45 & 0.94&\textbf{0.56}&\textbf{0.71} \\
      \midrule
      Calcite & 0.78&0.88&0.83 & \textbf{0.97}&\textbf{0.98}&\textbf{0.97}\\
      Ambari & 0.45&0.29&0.35 & \textbf{0.94}&\textbf{0.96}&\textbf{0.95}\\
      Cassandra & NA&NA&NA & \textbf{1.00}&\textbf{0.79}&\textbf{0.88}\\
      Freemarker & 0.39&0.88&0.54 & \textbf{0.73}&\textbf{1.00}&\textbf{0.85}\\
      \midrule
      Pig & 0.60&0.88&0.71 & \textbf{0.97}&\textbf{1.00}&\textbf{0.98}\\
      Derby & 0.40&0.82&0.53 & \textbf{0.96}&\textbf{0.99}&\textbf{0.98}\\
      Drools & 0.81&1.00&0.89 & \textbf{0.96}&\textbf{1.00}&\textbf{0.98}\\
      Flink & 0.54&\textbf{1.00}&0.70 & \textbf{0.93}&0.91&\textbf{0.92}\\
      \midrule
      Infinispan & 0.29&0.76&0.42 & \textbf{0.97}&\textbf{0.99}&\textbf{0.98}\\
      Arrow & 0.80&0.94&0.86 & \textbf{0.99}&\textbf{0.96}&\textbf{0.97}\\
      Airflow & 0.44&0.94&0.60 & \textbf{0.98}&\textbf{0.97}&\textbf{0.97}\\
      Causeway & 0.11&0.11&0.11 & \textbf{0.94}&\textbf{0.36}&\textbf{0.52}\\
      \midrule
      Ignite & 0.64&0.52&0.58 & \textbf{0.94}&\textbf{0.88}&\textbf{0.91}\\
      Groovy & 0.33&0.82&0.47 & \textbf{0.87}&\textbf{0.85}&\textbf{0.86}\\
      Pgcli & 0.46&\textbf{0.82}&0.59 & \textbf{0.75}&0.66&\textbf{0.70}\\
      Keras & \textbf{0.77}&\textbf{0.82}&\textbf{0.80} & 0.68&0.70&0.69\\
      \hline
      \rowcolor{babyblueeyes!70}
      Avg. & 0.59&0.73&0.61 & \textbf{0.90}&\textbf{0.87}&\textbf{0.88} \\
      \hline
      Standard deviation & 0.23&0.27&0.21 & \textbf{0.09}&\textbf{0.17}&\textbf{0.12} \\
     \bottomrule
    \end{tabular}
    \end{center}
    \vspace{-6mm}
\end{table}

Table~\ref{table:cross_project_results} displays the outcomes of the final experiment. Based on the average F1-measure over all folds, it is evident that LinkFormer outperforms the baseline by \checknum{$44$\%} in the cross-project mode as well.
In the cross-project setting, the models do not utilize any information from any of the projects in the test set (no historical data). Hence, it is anticipated that the average performance of both models would experience a decrease as we transition from the intermediate fine-tuning mode (RQ3) to the cross-project setting (RQ4). Our results confirm this hypothesis.
Similar to all three previous experiments, LinkFormer displays a lower standard deviation than the baseline model, T-BERT. Therefore, LinkFormer demonstrates a more consistent performance across all projects and the five folds.
Furthermore, LinkFormer achieves a balanced set of scores for precision and recall, while T-BERT obtains high recall and low precision in both fine-tuned and cross-project settings.
%
%
\begin{custombox}
\textbf{Insights based on the transfer learning setting}:
LinkFormer has demonstrated comparable performance in both intermediate fine-tuning and cross-project settings with the project-based mode (temporally-split datasets). These results illustrate the model's ability to transfer knowledge effectively from training projects to new and unseen test projects, making it a viable option for link prediction in situations where training data is limited or unavailable.
Hence, we recommend prioritizing the development of more generalizable models that can be effectively applied across a diverse range of software projects, while maintaining stable and consistent performance.
\end{custombox}

\section{Discussion}\label{discussion}
In this work, to tackle the link prediction task, we employ a combination of textual (i.e., text of issues and commit messages) and metadata (i.e., author, committer, timestamps, type, status, etc.) information of issue reports and commits to train our model. To more effectively leverage the contextual information in issues and commits and facilitate knowledge transfer for use on projects with insufficient data, we utilize pre-trained Transformer models.
Moreover, we investigate the impact of training-testing leakage and the generalizability of our proposed model to build more realistic and robust models.
Based on the results of our empirical assessment, we first discuss the findings and their implications. Subsequently, we provide a brief overview of the runtime characteristics of LinkFormer as a noteworthy feature of our study. Finally, we present several potential directions for future work.

\subsection{Splitting Data Policy and Data Leakage}
\label{sec:data_splitting_policy}

The concept of time has a significant impact on the link prediction problem for several reasons, including (1) the interactive nature of the life-cycle of open-source software development, and (2) the evolution and changing flow of a project.
Our findings indicate that time is indeed an important factor worthy of consideration when studying the $\langle issue, commit \rangle$ link prediction problem.
%
%
Random shuffling of datasets during model training can lead to inadvertent inclusion of future data in the learning process. This can result in model overfitting and decreased generalization performance. However, employing a temporal split of the data can effectively prevent the model from accessing future information during training.
The technique of masking future data is an effective solution to the problem of ``poking at future events" by machine learning models, as proposed by Zuo et al.~\cite{zuo2020transformer} and Zhang et al.~\cite{zhang2020self}. By masking future data, the model is forced to learn based on only the information available at each time step, thus preventing it from gaining an unfair advantage by accessing future information.
To our knowledge, the study by Rath et al.~\cite{rath2018traceability} is the only research work that has considered time in the context of link prediction among issues and commits. In their study, the authors incorporated time as an additional feature during model training. However, they did not investigate the effects of various data splitting techniques, nor did they evaluate all possible types of issues. Specifically, their analysis was limited to bug- and implementation-related issues only.
Moreover, they use a classical J48 algorithm
to train multiple project-based classifiers.
In this study, we experiment 
with different splitting settings 
and empirically show their impact.
We also train more accurate and generalizable models 
using deep pre-trained models 
to improve upon generalizability.
Hence, we only need a single model 
to recover links for different types of issues, 
and for different projects.

The results of RQ1 and RQ2 demonstrate the impact of the choice of data splitting policies (random and temporal) in the case of link prediction models. Although the behavior varies per project as well as per model, on average, all approaches are negatively affected when moving from a random to a temporal splitting policy. These findings suggest that previous studies have overestimated the performance of their models when using randomly-split data.
Note that the performance of different models varies in their sensitivity to the choice of data splitting policies. T-BERT is drastically affected, with a $25\%$ drop in the average F1-measure when moving from random to temporal policy, while LinkFormer is the least affected model, with only a $7\%$ reduction in the average F1-measure.
In the future work, we aim to investigate the reasons for such varying impact.

\begin{custombox}
\textbf{Recommendation:}
Given the time-aware nature of the problem, this study strongly recommends the adoption of temporal splitting policy to assess the performance of learning-based models in real-world settings. The temporal splitting policy ensures that no future data is inadvertently included in the training process, which can lead to biased model performance and unrealistic capabilities. Therefore, by using the temporal splitting policy, researchers can simulate the performance of learning-based models in a more realistic manner and make informed decisions regarding the application of these models in practice.
\end{custombox}

\subsection{Generalizabilty}
For a model to be considered generalizable, it should be capable of predicting missing links, even when applied to new projects with limited available data. To achieve this goal, we developed a generic LinkFormer model and trained it using both intermediate fine-tuning and cross-project settings. 
This approach enabled the model to acquire knowledge from existing software projects and transfer it to unseen ones. As a result, the model can effectively predict missing links in new projects, even in situations where the available data is limited.
Based on the results obtained from our four experiments conducted in various settings, we can conclude that LinkFormer's performance in the transfer learning modes (i.e., fine-tuning and cross-project) is comparable to that of the project-based mode (i.e., temporal split). This indicates that the model can effectively transfer knowledge across different software projects without significant degradation in performance.
Additionally, the low standard deviation observed across all experiments further strengthens the robustness of the proposed approach. LinkFormer's low standard deviation indicates that the model's performance is stable across all projects and settings, which suggests that the proposed approach is highly generalizable and can be applied to predict missing links for any unseen project, even with limited or no available data.
T-BERT, the state-of-the-art model pre-trained initially on the CodeSearchNet dataset and then fine-tuned on our dataset, also exhibits comparable performance in both transfer and project-based modes. However, the accuracy of T-BERT, in both fine-tuned and cross-project settings, is much lower compared to LinkFormer, and it also exhibits greater fluctuations across projects. 

\begin{custombox}
\textbf{Recommendation:}
Given the potential benefits of pre-trained models and the unique characteristics of the link prediction problem, we strongly recommend that researchers invest in developing more generalizable models that can be applied to various types of software projects with stable and consistent performance. 
By focusing on the development of more generalizable models, researchers can enhance the efficiency and accuracy of software development processes, reduce the time and cost associated with manual traceability tasks, and enable the automation of complex software engineering tasks.
\end{custombox}


\subsection{Runtime Characteristics and Underlying Architectures}
In addition to accuracy, the practical considerations of runtime efficiency are also crucial for the widespread adoption of predictive models in software engineering. Fortunately, LinkFormer provides both accurate predictions and good runtime characteristics, making it a practical and effective solution for software traceability tasks. The project-based mode of LinkFormer 
can be trained on average in just 8 minutes per project, which is significantly faster than training a generic model using batches of projects altogether in the cross-project setting with the same architecture, which takes 43 minutes to fully train. This is because the generic model is trained on more data. However, it is important to note that training time is a one-time cost. Once the model is trained, its inference time is quite low, with the entire test set of the Flink project taking only 12 seconds to complete predictions. Overall, LinkFormer provides a practical and efficient solution for software traceability tasks, with the potential to significantly improve the productivity and accuracy of software development processes.

The choice of architecture is also an important factor to consider when building link prediction models.
In our study, we evaluated the performance of LinkFormer using three different transformer architectures, namely, RoBERTa, DistilBERT, and AlBERT, to better understand the impact of these different models on the link prediction task in software engineering.
Our investigation of these architectures in building LinkFormer shows that the fine-tuned version based on the pre-trained RoBERTa model generates the most accurate predictions. However, in some cases, the DistilBERT or AlBERT models can also achieve high F1-measure values. 
For the details of the results, please refer to Appendix~\ref{appendix:arch}.

Lastly, it is worth mentioning that in this work, inspired by the challenges encountered while trying to reuse other models as baselines, which often involved complex setup and configuration, we aimed to provide a simple and easy-to-use model to achieve broader adoption and facilitate the reuse of our approach by other researchers and software developers.

\begin{custombox}
\textbf{Recommendation:}
The runtime efficiency of predictive models is a critical factor for their widespread adoption in the software industry. Researchers and practitioners should consider the practical aspects of integrating these models into existing software development workflows. Ultimately, this will enable software organizations to develop high-quality software products faster and more reliably.
Additionally, the underlying architecture is another factor when building such predictive models.
How to select the best architecture ultimately depends on the specific needs of the software development process, taking into account factors such as model size, training and inference time, and the desired level of accuracy.
For example, smaller architectures such as DistilBERT and AlBERT can be preferred when computational resources are limited, while larger architectures such as RoBERTa can be preferred when higher accuracy is desired.
\end{custombox}

\subsection{Future Direction}
In this work, we focused on 
improving the accuracy of link recovery techniques 
among issues and commits in collaborative software development.
Moreover, we considered different angles of this problem 
including temporal aspects of data 
and generalizability of the models.
As we primary focus on one-to-one links among issues and commits, 
extending the current work to include many-to-many links among issues and commits would be an interesting avenue for future research. 
This could involve exploring more complex models, such as graph neural networks, that are better suited for modeling relationships between multiple entities. 
An area that warrants further exploration is the development of more appropriate approaches for processing source code and extracting pertinent information, with the aim of achieving more precise linking in cross-project contexts, all while keeping the computational overhead of the model to a minimum.

Furthermore, our current approach involves the construction and utilization of well-balanced datasets to effectively train our machine learning models. It is feasible to implement balancing techniques directly within the learning model, thereby enabling us to thoroughly examine the impact of various balancing techniques. 

Application of LinkFormer to other software artifacts, including but not limited to test cases, requirements documentation, and so forth is also an interesting avenue to pursue. Moreover, it would be worthwhile to examine the influence of utilizing distinct data sources, such as code reviews or pull requests, on the precision of link recovery techniques.

Lastly, there is potential to further enhance the negative sampling technique employed in our approach. In a study conducted by Grammel et al.~\cite{grammel2010attracting}, the involvement of the community in IBM Jazz closed-source projects was investigated. Their findings indicated that issues created by community members may take longer to address than those created by project members. While our models consider the role of issue openers, we utilize a fixed-length window for negative sampling. Hence, it would be valuable for future research to propose novel negative sampling techniques that account for this discrepancy and improve the performance of predictive models.



\subsection{Threats to Validity}\label{threat}
In the ensuing section, we explicate the potential threats to the validity of this research, 
classified as internal, external, and construct threats.

\paragraph{Internal Validity:}
Internal validity refers to the degree to which a given evidence substantiates a claim concerning the causal relationship between variables, in the specific context of a particular study~\cite{ampatzoglou2019identifying}.
The primary threat to the internal validity of our study pertains to the reliability of the True Links and False Links utilized in our research. In order to minimize this potential threat, we employed established datasets that are commonly utilized in the relevant literature~\cite{mazrae2021automated,rath2018traceability,lin2021traceability}.
This practice can enhance the credibility and robustness of research findings, and facilitate the replication and comparison of results across studies.
Furthermore, given the vast number of potential False Links,  
we relied on the heuristics proposed by previous work to constrain the number of False Links in our study~\cite{mazrae2021automated,sun2017frlink}.
Finally, while creating a balanced dataset for projects,
we randomly selected from the set of False Links 
to avoid introducing selection bias.
Although each of these datasets is validated by other researchers~\cite{rath2018traceability,lin2021traceability,mazrae2021automated}, 
incorrect links may still be present due to human error.
Time also plays a role 
when generating the False Links using a negative sampling technique.
Each project has several date fields
which are used for
constructing the final sets for training and testing. 
Different studies use a series of constraints 
on the aforementioned date fields 
to shrink the space of potential False Links.
The literature~\cite{ruan2019deeplink,mazrae2021automated,lin2021traceability} 
introduces various time constraints for commit dates. 
Mostly, the literature assumes~\cite{mazrae2021automated,ruan2019deeplink,sun2017frlink} 
if an issue and a commit are related, 
they should have close submission dates (within one week).
Hence, we used the same rule.
However, there exist links that are more than seven days apart. 
Future research should devise better negative sampling techniques.
%

\paragraph{External Validity:} 
External validity is concerned with the generalizability 
of the approach and results~\cite{ampatzoglou2019identifying}.
To increase the generalizability of our approach, 
we collected and assessed the performance of our approach
on a set of $20$ projects collected
from three different studies in the field.
The project goal and type vary 
from data processing frameworks
to programming languages, 
deep learning APIs, and many more.
Moreover, as different platforms 
provide different linking behavior,
we included two combinations of
issue tracking system (Jira) and version control system (Git)
among the projects.
The main programming language 
also varies for the selected projects.
\minor{Note that LinkFormer achieves the lowest standard deviation in all settings
indicating that its results are more stable across projects.}
As we focused on open-source projects,
there is a potential threat that 
our findings can not be generalized to commercial projects.
Although there are similarities in open source and commercial projects,
companies' policies may affect the commit practices.
The impact of such decisions 
should be investigated in future research.
\minor{To evaluate our generic model more fairly, 
we validate it using five folds and report both individual and the average scores.
By breaking data into five smaller chunks and re-evaluating the model, 
we ensure that all of the data has been used for training and testing.
Finally, our best model outperforms the baselines 
in at least 19 projects out of 20 projects in all settings.
However, LinkFormer's average scores 
are significantly higher in all settings 
than those of the baselines.
}

\paragraph{Construct Validity:}
Construct validity is concerned 
with the evaluation of the models~\cite{ampatzoglou2019identifying}.
Similar to previous work~\cite{ruan2019deeplink,sun2017frlink,sun2017improving,schermann2015discovering,le2015rclinker,rath2018traceability,mazrae2021automated,lin2021traceability,izadi2022evaluation}, 
we use the standard metric F1-measure  
commonly used in the literature 
as the harmonic mean of precision and recall scores
to evaluate the performance of our approach 
against the state-of-the-art.
We also report precision and recall of these approaches.


\section{Conclusion}\label{conclusion}
The automated linking of commits to their corresponding issue reports is a widely used practice in software engineering to enhance various software maintenance tasks. These tasks include software documentation, defect prediction, bug localization, software quality measurement, and many more. 
This research presents LinkFormer to address crucial challenges related to the link recovery problem, particularly those related to data leakage and the generalizability of proposed solutions. 
We provide empirical evidence demonstrating the impact of train-test set splitting policies on the performance of models used in this context. Additionally, we train and evaluate two generic models under two transfer-learning settings, and show that our proposed approach performs comparably to the project-based setting. 

The outcomes of our research hold implications for the development of effective link recovery methods and can inform software engineering practices pertaining to software maintenance and development. Specifically, we recommend the utilization of a temporal splitting policy to evaluate the performance of learning-based models in real-world settings, as this approach is more reflective of practical scenarios. Furthermore, we posit that investing in the development of more generalizable models that can be applied across various types of software projects and exhibit stable and consistent performance is critical in this context. 
Future research can build on this work by focusing on the development of generalizable models that can effectively process source code in addition to other relevant data sources. This would enable the construction of more comprehensive link prediction systems that accommodate a wider range of software development scenarios and ultimately lead to more effective software maintenance and development practices.

\begin{acknowledgements}
This work is partially supported by the ARC-21/25 UMONS3 Action de Recherche Concertée financée par le Ministère de la Communauté française – Direction générale de l’Enseignement non obligatoire et de la Recherche scientifique.
This work is also supported by
DigitalWallonia4.AI research project ARIAC (grant number 2010235).
\end{acknowledgements}

\section*{Conflict of Interest}\label{COI}
The authors have no conflict of interest.

\section*{Data Availability Statement}\label{DAS}
The dataset used for training the models and comparing approaches 
is publicly available online.~\footnote{\url{https://doi.org/10.5281/zenodo.6524460}}

\bibliographystyle{abbrv} 
\bibliography{main}
%
\begin{appendices}

\section{Precision and Recall scores}
\label{appendix:precision_recall}

As discussed in Section~\ref{sec:eval_metrics},
a good link prediction method 
shall have both high precision and recall.
Tables~\ref{table:random_split_results_appendix} and \ref{table:temporal_split_results_appendix} contain the result of LinkFormer in comparison with state-of-the-art in a detailed manner in case of the train and test data split in a random and temporal manner.
All the baselines and LinkFormer's predictions 
obtain a balanced precision and recall in the random splitting setting.
However, when they are tested on temporally-split data,
their performance becomes unbalanced.
For instance, T-BERT's precision drops more than its recall.
On the contrary, DeepLink's recall decreases more than its precision, 
resulting in an unbalanced performance for this model.
HybridLinker and LinkFormer 
both maintain a balanced precision and recall.

\begin{table*}[h]
    \caption{Project-based training and assessment, full details (RQ1: randomly-split datasets)}
    \begin{center}
     \begin{tabular}{c|cca|cca|cca|cca} 
     \toprule
     \multirow{2}{*}{\textbf{Project}} 
     & \multicolumn{3}{c}{Deeplink~\cite{ruan2019deeplink}} & \multicolumn{3}{c}{Hybrid-Linker~\cite{mazrae2021automated}} & \multicolumn{3}{c}{T-BERT~\cite{lin2021traceability}} & \multicolumn{3}{c}{LinkFormer (Proposed)}\\\cmidrule{2-13}
     & P & R & F1 
     & P & R & F1 
     & P & R & F1
     & P & R & F1\\[0.15ex]
     \midrule
      Maven      & 0.69&0.64&0.66 & 0.79&0.82&0.81  & 0.93&0.88&0.90 & \textbf{1.00}&\textbf{1.00}&\textbf{1.00} \\
      Pig        & 0.65&0.68&0.66 & 0.71&0.76&0.74  & 1.00&1.00&1.00 & \textbf{1.00}&\textbf{1.00}&\textbf{1.00} \\
      Derby      & 0.57&0.70&0.63 & 0.87&0.83&0.85  & 1.00&1.00&1.00 & \textbf{1.00}&\textbf{1.00}&\textbf{1.00} \\
      Drools     & 0.65&0.64&0.64 & 0.84&0.84&0.84  & 1.00&1.00&1.00 & \textbf{1.00}&\textbf{1.00}&\textbf{1.00} \\
      Infinispan & 0.57&0.54&0.55 & 0.86&0.89&0.87  & 1.00&1.00&1.00 & \textbf{1.00}&\textbf{1.00}&\textbf{1.00} \\
      \hline
      Cassandra & 0.84&0.76&0.80 & 0.84&1.00&0.91   & NA&NA&NA & \textbf{0.93}&0.87&0.90 \\
      Freemarker& 0.95&0.84&0.89 & 0.86&0.88&0.87  & 0.06&0.94&0.11 & \textbf{0.95}&\textbf{1.00}&\textbf{0.97} \\
      Netbeans  & 0.84&0.45&0.59 & 0.87&0.89&0.88  & 0.56&0.76&0.65 & \textbf{0.96}&\textbf{0.92}&\textbf{0.94} \\
      Calcite   & 0.67&0.66&0.66 & 0.86&0.89&0.87  & 0.94&0.94&0.94 & \textbf{0.99}&\textbf{0.99}&\textbf{0.99} \\
      Arrow     & 0.47&0.40&0.43 & 0.84&0.84&0.84  & 0.94&0.94&0.94 & \textbf{0.97}&\textbf{0.98}&\textbf{0.97} \\ 
      Airflow   & 0.70&0.33&0.45 & 0.85&0.87&0.86  & 0.94&\textbf{1.00}&0.97 & \textbf{0.99}&0.97&\textbf{0.98} \\ 
      Beam      & 0.70&0.62&0.66 & 0.86&0.86&0.86  & 0.61&0.82&0.70 & \textbf{0.95}&\textbf{0.96}&\textbf{0.95} \\
      Causeway      & 0.74&0.85&0.79 & 0.87&0.88&0.88  & \textbf{0.99}&0.88&0.92 & 0.92&\textbf{0.92}&\textbf{0.92} \\ 
      Groovy    & 0.62&0.66&0.64 & 0.88&0.89&0.88  & \textbf{1.00}&0.94&\textbf{0.97} & 0.96&\textbf{0.96}&0.96 \\
      Ignite    & 0.63&0.66&0.64 & 0.89&0.91&0.90  & 0.75&0.88&0.81 & \textbf{0.94}&\textbf{0.94}&\textbf{0.94} \\
      Flink     & 0.67&0.78&0.72 & 0.90&0.92&0.91  & 0.94&0.94&0.94 & \textbf{0.98}&\textbf{0.96}&\textbf{0.97} \\
      Ambari    & 0.83&0.81&0.82 & 0.95&0.97&0.96  & \textbf{1.00}&\textbf{1.00}&\textbf{1.00} & 0.99&0.99&0.99 \\
      \hline
      Pgcli & 0.53&0.45&0.49 & 0.61&0.75&0.67  & 0.81&\textbf{1.00}&0.89 & \textbf{0.97}&0.92&\textbf{0.95} \\
      Keras & 0.43&0.30&0.35 & 0.78&0.72&0.75  & \textbf{1.00}&0.94&0.97 & 0.99&\textbf{1.00}&\textbf{0.99} \\
      Flask & 0.48&0.37&0.42 & 0.66&0.69&0.68  & \textbf{1.00}&0.58&0.74 & 0.91&\textbf{0.91}&\textbf{0.91} \\
      \hline
      \rowcolor{babyblueeyes!70}
      Average & 0.66&0.60&0.62 & 0.83&0.85&0.84 &  0.86&0.91&0.86 & \textbf{0.97}&\textbf{0.96}&\textbf{0.97}\\
      \hline
      Standard deviation & 0.13&0.17&0.15 & 0.08&0.07&0.07 &  0.24&0.11&0.21 & \textbf{0.02}&\textbf{0.03}&\textbf{0.03}\\
     \bottomrule
    \end{tabular}
    \end{center}
    \label{table:random_split_results_appendix}
\end{table*}

\begin{table*}[h]
    \caption{Project-based training and assessment, full details (RQ2:  temporally-split datasets)}
    \begin{center}
     \begin{tabular}{c|cca|cca|cca|cca} 
     \toprule
     \multirow{2}{*}{\textbf{ID}}
      & \multicolumn{3}{c}{Deeplink~\cite{ruan2019deeplink}} & \multicolumn{3}{c}{Hybrid-Linker~\cite{mazrae2021automated}} & \multicolumn{3}{c}{T-BERT~\cite{lin2021traceability}} & \multicolumn{3}{c}{LinkFormer (Proposed)}\\\cmidrule{2-13}
     & P & R & F1 
     & P & R & F1 
     & P & R & F1
     & P & R & F1\\[0.15ex]
     \midrule
      Maven        & 0.74&0.83&0.79 & 0.74&0.99&0.85& 0.71&0.88&0.78 & \textbf{0.95}&\textbf{1.00}&\textbf{0.97} \\
      Pig          & 0.87&0.79&0.83 & 0.64&0.94&0.76& 0.60&1.00&0.75 & \textbf{0.98}&\textbf{1.00}&\textbf{0.99} \\
      Derby        & 0.69&0.35&0.47 & 0.79&0.94&0.86& 0.38&0.94&0.54 & \textbf{1.00}&\textbf{1.00}&\textbf{1.00} \\
      Drools       & 0.71&0.35&0.47 & 0.69&0.79&0.74& 0.94&0.94&0.94 & \textbf{0.99}&\textbf{1.00}&\textbf{0.99} \\
      Infinispan   & 0.67&0.50&0.57 & 0.78&0.93&0.85& 0.30&0.76&0.43 & \textbf{0.98}&\textbf{1.00}&\textbf{0.99} \\
      \hline
      Cassandra  & 0.81&0.69&0.75 & 0.67&0.59&0.63& NA&NA&NA & \textbf{0.83}&\textbf{0.71}&\textbf{0.77} \\
      Freemarker & 0.12&0.14&0.13 & 0.28&\textbf{0.80}&0.42 & 0.10&0.35&0.16 & \textbf{0.79}&0.65&\textbf{0.71} \\
      Netbeans   & 0.30&0.05&0.09 & 0.42&0.46&0.44& 0.86&0.76&0.81 & \textbf{0.89}&\textbf{0.92}&\textbf{0.90} \\
      Calcite    & 0.59&0.35&0.44 & 0.67&0.55&0.60& 0.88&0.94&0.91 & \textbf{0.98}&\textbf{0.96}&\textbf{0.97} \\
      Arrow      & 0.65&0.32&0.43 & 0.79&0.62&0.69& 0.88&0.94&0.91 & \textbf{0.98}&\textbf{0.96}&\textbf{0.97} \\
      Airflow    & 0.66&0.62&0.64 & 0.67&0.45&0.54& 0.45&0.88&0.60 & \textbf{0.99}&\textbf{0.96}&\textbf{0.98} \\
      Beam       & 0.68&0.44&0.54 & 0.74&0.47&0.58& 0.63&0.41&0.50 & \textbf{0.95}&\textbf{0.62}&\textbf{0.75} \\
      Causeway       & 0.70&0.43&0.54 & 0.74&0.27&0.40& 0.13&0.11&0.12 & \textbf{0.82}&\textbf{0.61}&\textbf{0.70} \\
      Groovy     & 0.75&0.50&0.60 & 0.70&0.44&0.54& 0.33&\textbf{1.00}&0.50 & \textbf{0.95}&0.87&\textbf{0.91} \\
      Ignite     & 0.59&0.47&0.52 & 0.85&0.73&0.78& 0.37&0.88&0.52 & \textbf{0.96}&\textbf{0.92}&\textbf{0.94} \\
      Flink      & 0.65&0.65&0.65 & 0.85&0.56&0.68& 0.84&0.64&0.73 & \textbf{0.99}&\textbf{0.89}&\textbf{0.94} \\
      Ambari     & 0.71&0.63&0.67 & 0.93&0.77&0.84& 0.23&0.94&0.37 & \textbf{1.00}&\textbf{0.89}&\textbf{0.94} \\
      \hline
      Pgcli  & 0.56&0.68&0.62 & 0.57&\textbf{0.99}&0.72& 0.70&0.70&0.70 & \textbf{0.76}&0.73&\textbf{0.75} \\
      Keras  & 0.54&0.47&0.50 & 0.64&\textbf{1.00}&0.78 & 0.82&0.82&0.82 & \textbf{0.89}&0.90&\textbf{0.90} \\
      Flask  & 0.42&0.86&0.57 & 0.48&0.88&0.62& 0.85&0.35&0.50 & \textbf{0.95}&\textbf{0.91}&\textbf{0.93} \\
      \hline
      \rowcolor{babyblueeyes!70}
      Avg.  & 0.62&0.50&0.54 & 0.68&0.71&0.67 & 0.57&0.74&0.61 & \textbf{0.93}&\textbf{0.88}&\textbf{0.90}\\
      \hline
      Standard deviation   & 0.17&0.21&0.18 & 0.15&0.22&0.14 & 0.28&0.26&0.23 & \textbf{0.07}&\textbf{0.13}&\textbf{0.10}\\
     \bottomrule
    \end{tabular}
    \end{center}
    \label{table:temporal_split_results_appendix}
\end{table*}

\section{Comparison among Underlying Architectures for LinkFormer}
\label{appendix:arch}
Tables~\ref{table:random_split_BERT_submodels_comparison} compares our proposed model in comparison to two other models in randomly-split data mode, while Table~\ref{table:temporal_split_BERT_submodels_comparison} shows the results for the temporally-split mode.
Moreover, Table~\ref{table:fine_tuning_results_BERT_submodels_comparison} compares the performance of these models in case of intermediate fine-tuning (RQ3) and Table~\ref{table:cross_project_results_BERT_submodels_comparison} provides the experminets results in the case of cross-project.

\begin{table*}[tb]
    \caption{Project-based training and assessment of Transformer-based architectures (RQ1: randomly-split datasets)}
    \begin{center}
     \begin{tabular}{l|r|r|r} 
     \toprule
     \textbf{Project} & LinkFormer & DistilBERT & AlBERT\\[0.15ex]
     \midrule
       Maven   & \textbf{1.00} & \textbf{1.00} & \textbf{1.00} \\
       Pig    & \textbf{1.00} & \textbf{1.00} & \textbf{1.00} \\
       Derby   & \textbf{1.00} & \textbf{1.00} & \textbf{1.00} \\
       Drools   & \textbf{1.00} & \textbf{1.00} & \textbf{1.00} \\
       Infinispan   & \textbf{1.00} & \textbf{1.00} & \textbf{1.00} \\
       \hline
       Cassandra    & \textbf{0.90} & 0.85 & 0.89 \\
       Freemarker    & \textbf{0.97} & \textbf{0.97} & \textbf{0.97} \\
       Netbeans    & \textbf{0.94} & \textbf{0.94} & 0.93 \\
       Calcite    & \textbf{0.99} & 0.98 & 0.98 \\
       Arrow   & \textbf{0.97} & \textbf{0.97} & \textbf{0.97} \\ 
       Airflow  & \textbf{0.98} & \textbf{0.98} & \textbf{0.98} \\ 
       Beam   & \textbf{0.95} & 0.94 & 0.93 \\
       Causeway   & \textbf{0.92} & \textbf{0.92} & 0.91 \\ 
       Groovy  & \textbf{0.96} & 0.94 & 0.93 \\
       Ignite   & \textbf{0.94} & \textbf{0.94} & 0.93 \\
       Flink   & \textbf{0.97} & 0.96 & 0.95 \\
       Ambari  & \textbf{0.99} & \textbf{0.99} & \textbf{0.99} \\
       \hline
       Pgcli   & \textbf{0.95} & 0.93 & 0.92 \\
       Keras   & \textbf{0.99} & 0.98 & \textbf{0.99} \\
       Flask   & 0.91 & 0.91 & \textbf{0.93} \\
       \hline
       \rowcolor{babyblueeyes!70}
       Average score & \textbf{0.97} & 0.96 & 0.96 \\
       \hline
       Standard deviation & \textbf{0.03} & \textbf{0.03} & \textbf{0.03} \\
     \bottomrule
    \end{tabular}
    \end{center}
    \label{table:random_split_BERT_submodels_comparison}
    \vspace{-6mm}
\end{table*}

\begin{table*}[tb] 
    \caption{Project-based training and assessment of Transformer-based architectures (RQ2: temporally-split datasets}
    \begin{center}
     \begin{tabular}{l|r|r|r}
     \toprule
     \textbf{Project} & LinkFormer & DistilBERT & AlBERT\\[0.15ex]
     \midrule
       Maven   & 0.97 & 0.94 & \textbf{1.00} \\
       Pig   & 0.99 & 0.99 & \textbf{1.00} \\
       Derby   & \textbf{1.00} & 0.99 & \textbf{1.00} \\
       Drools  & 0.99 & 0.99 & \textbf{1.00} \\
       Infinispan   & \textbf{0.99} & \textbf{0.99} & \textbf{0.99} \\
       \hline
       Cassandra  & 0.77 & \textbf{0.86} & 0.35\\
       Freemarker  & \textbf{0.71} & 0.57 & 0.64\\
       Netbeans   & \textbf{0.90} & \textbf{0.90} & \textbf{0.90}\\
       Calcite    & 0.97 & 0.97 & \textbf{0.98}\\
       Arrow   & \textbf{0.97} & \textbf{0.97} & \textbf{0.97}\\
       Airflow   & \textbf{0.98} & \textbf{0.98} & 0.97\\
       Beam   & 0.75 & 0.69 & \textbf{0.88}\\
       Causeway   & \textbf{0.70} & 0.60 & 0.69\\
       Groovy   & \textbf{0.91} & 0.88 & 0.87\\
       Ignite   & \textbf{0.94} & 0.92 & 0.92\\
       Flink   & \textbf{0.94} & 0.92 & 0.93\\
       Ambari   & 0.94 & \textbf{0.95} & 0.93\\
       \hline
       Pgcli   & 0.75 & 0.80 & \textbf{0.82}\\
       Keras   & \textbf{0.90} & 0.89 & 0.89\\
       Flask   & \textbf{0.93} & \textbf{0.93} & 0.91\\
       \hline
       \rowcolor{babyblueeyes!70}
       Average score & \textbf{0.90} & 0.89 & 0.88\\
       \hline
       Standard deviation & \textbf{0.10} & 0.12 & 0.15\\
     \bottomrule
    \end{tabular}
    \end{center}
    \label{table:temporal_split_BERT_submodels_comparison}
    \vspace{-6mm}
\end{table*}

\begin{table}[tb]
    \caption{Transfer learning on Transformer-based architectures (RQ3: intermediate fine-tuning)}
    \begin{center}
     \begin{tabular}{l|r|r|r} 
     \toprule
     \textbf{Project} & LinkFormer & DistilBERT & AlBERT \\[0.15ex]
     \midrule
       Maven & 0.99&\textbf{1.00}&0.99\\
       Netbeans & 0.92&\textbf{0.93}&0.90\\
       Flask  & 0.90&\textbf{0.92}&0.87\\
       Beam & \textbf{0.75}&0.67&0.70\\
       \midrule
       Calcite & \textbf{0.97}&0.96&0.96\\
       Ambari & \textbf{0.95}&\textbf{0.95}&\textbf{0.95}\\
       Cassandra & \textbf{0.88}&0.80&0.85\\
       Freemarker & 0.74&\textbf{0.78}&0.74\\
       \midrule
       Derby & \textbf{0.99}&0.97&0.72\\
       Drools & \textbf{0.99}&\textbf{0.99}&0.85\\
       Pig & \textbf{0.99}&0.95&0.77\\
       Flink & \textbf{0.93}&0.90&0.87\\
       \midrule
       Infinispan & \textbf{0.99}&\textbf{0.99}&0.97\\
       Arrow & \textbf{0.97}&0.94&0.94\\
       Airflow & \textbf{0.95}&0.93&0.93\\
       Causeway & 0.70&\textbf{0.72}&0.63\\
       \midrule
       Ignite & \textbf{0.92}&0.90&\textbf{0.92}\\
       Groovy & \textbf{0.88}&0.83&0.87\\
       Keras & \textbf{0.83}&0.75&0.69\\
       Pgcli & \textbf{0.82}&0.76&0.74\\
       \hline
       \rowcolor{babyblueeyes!70}
       Average score & \textbf{0.90}&0.88&0.84\\
       \hline
       Standard deviation & \textbf{0.09}&0.10&0.10\\
     \bottomrule
    \end{tabular}
    \end{center}
    \label{table:fine_tuning_results_BERT_submodels_comparison}
\end{table}

\begin{table}[tb]
    \caption{Transfer learning on Transformer-based architectures (RQ4: cross-project mode)}
    \label{table:cross_project_results_BERT_submodels_comparison}
    \begin{center}
     \begin{tabular}{l|r|r|r} 
     \toprule
     \textbf{Project} & LinkFormer & DistilBERT & AlBERT \\[0.15ex]
     \midrule
      Maven & \textbf{0.99}&\textbf{0.99}&0.96 \\
      Netbeans & \textbf{0.91}&0.88&0.89 \\
      Flask  & 0.84&\textbf{0.88}&0.76 \\
      Beam & \textbf{0.71}&0.65&0.65 \\
      \midrule
      Calcite & \textbf{0.97}&\textbf{0.97}&\textbf{0.97} \\
      Ambari & 0.95&\textbf{0.96}&\textbf{0.96} \\
      Cassandra & \textbf{0.88}&0.83&0.80 \\
      Freemarker & \textbf{0.85}&0.80&0.74 \\
      \midrule
      Pig & \textbf{0.98}&0.96&0.94 \\
      Derby & \textbf{0.98}&0.97&0.89 \\
      Drools & \textbf{0.98}&0.97&0.97 \\
      Flink & \textbf{0.92}&0.91&0.90 \\
      \midrule
      Infinispan & \textbf{0.98}&\textbf{0.98}&\textbf{0.98} \\
      Arrow & \textbf{0.97}&0.96&\textbf{0.97} \\
      Airflow & \textbf{0.97}&0.95&0.96 \\
      Causeway & 0.52&\textbf{0.65}&0.63 \\
      \midrule
      Ignite & 0.91&0.91&\textbf{0.93} \\
      Groovy & \textbf{0.86}&0.85&0.83 \\
      Pgcli & 0.70&\textbf{0.77}&0.75 \\
      Keras & 0.69&\textbf{0.76}&0.75 \\
      \hline
      \rowcolor{babyblueeyes!70}
      Average score & \textbf{0.88}&\textbf{0.88}&0.86 \\
      \hline
      Standard deviation & 0.12&\textbf{0.10}&0.11 \\
     \bottomrule
    \end{tabular}
    \end{center}
\end{table}

\end{appendices}

\end{document}